\documentclass[reprint,amsmath,superscriptaddress,amssymb,aps,prb]{revtex4-1}
\usepackage{graphicx}
\usepackage{dcolumn}
\usepackage{bm}
\usepackage{color,graphicx,mfirstuc}
\usepackage[colorlinks,linkcolor=blue,
            citecolor=blue,
            hyperindex,
            pdfstartview=FitH,
            plainpages=false]
            {hyperref}

\begin{document}

\title{Electron-phonon Coupling and Kohn Anomaly due to the Floating 2D Electronic Bands on the Surface of ZrSiS}

\author{Siwei Xue}
\affiliation{Beijing National Laboratory for Condensed Matter
Physics and Institute of Physics, Chinese Academy of Sciences,
Beijing 100190, China}
\affiliation{School of Physical
Sciences, University of Chinese Academy of Sciences, Beijing 100049,
China}
\author{Tiantian Zhang}
\affiliation{Beijing National Laboratory for Condensed Matter
Physics and Institute of Physics, Chinese Academy of Sciences,
Beijing 100190, China}
\author{Changjiang Yi}
\author{Shuyuan Zhang}
\author{Xun Jia}
\affiliation{Beijing National Laboratory for Condensed Matter
Physics and Institute of Physics, Chinese Academy of Sciences,
Beijing 100190, China}
\affiliation{School of Physical
Sciences, University of Chinese Academy of Sciences, Beijing 100049,
China}
\author{Luiz H. Santos}
\affiliation{Department of Physics, Emory University, 400 Dowman Drive, Atlanta, Georgia 30322, USA}
\author{Chen Fang}
\affiliation{Beijing National Laboratory for
Condensed Matter Physics and Institute of Physics, Chinese Academy
of Sciences, Beijing 100190, China}
\affiliation{Songshan Lake
Materials Laboratory, Dongguan, Guangdong 523808, China}
\affiliation{CAS Center for Excellence in Topological Quantum Computation,
University of Chinese Academy of Sciences, Beijing 100190, China.}
\author{Youguo Shi}
\affiliation{Beijing National Laboratory for
Condensed Matter Physics and Institute of Physics, Chinese Academy
of Sciences, Beijing 100190, China}
\affiliation{Songshan Lake
Materials Laboratory, Dongguan, Guangdong 523808, China}
\author{Xuetao Zhu}
\email{xtzhu@iphy.ac.cn}
\affiliation{Beijing National Laboratory for
Condensed Matter Physics and Institute of Physics, Chinese Academy
of Sciences, Beijing 100190, China}
\affiliation{School
of Physical Sciences, University of Chinese Academy of Sciences,
Beijing 100049, China}
\affiliation{Songshan Lake
Materials Laboratory, Dongguan, Guangdong 523808, China}
\author{Jiandong Guo}
\email{jdguo@iphy.ac.cn}
\affiliation{Beijing National Laboratory
for Condensed Matter Physics and Institute of Physics, Chinese
Academy of Sciences, Beijing 100190, China}
\affiliation{School of Physical Sciences, University of Chinese Academy of
Sciences, Beijing 100049, China}
\affiliation{Songshan
Lake Materials Laboratory, Dongguan, Guangdong 523808,
China}
\affiliation{Beijing Acadamy of Quantum
Information Sciences, Beijing 100193, China}

\date{\today}

\begin{abstract}
${\rm ZrSiS}$, an intriguing candidate of topological nodal line semimetals, was discovered to have exotic surface floating two-dimensional (2D) electrons [Phys. Rev. X 7, 041073 (2017)], which are likely to interact with surface phonons. Here, we reveal a prominent Kohn anomaly in a surface phonon branch by mapping out the surface phonon dispersions of ${\rm ZrSiS}$ using high-resolution electron energy loss spectroscopy. Theoretical analysis via an electron-phonon coupling (EPC) model attributes the strong renormalization of the surface phonon branch to the interactions with the surface floating 2D electrons. With the random phase approximation, we calculate the phonon self-energy and evaluate the mode-specific EPC constant by fitting the experimental data. The EPC picture provided here may be important for potential applications of topological nodal line semimetals.
\end{abstract}

\maketitle

\section{\label{sec:intro}Introduction}
Lattice and charge are definitely the two most essential degrees of freedom in condensed matter. The interplay between them, {\it i.e.}, the electron-phonon coupling (EPC), as a prototypical many-body problem, has received intensive theoretical and experimental investigations \cite{Grimvall.1981,Maksimov.1997,Pavarini.2017,Giustino.2017}. EPC is exceptionally important not only because it is closely bound up with macroscopic physical properties such as charge carrier dynamics, thermal conductivity, heat capacity and so on, but also due to its crucial roles in the microscopic mechanisms of many notable phenomena such as polarons, superconductivity, and charge density waves, {\it etc.}.

EPC changes the dispersion and lifetime of both the involving electrons and phonons. The effect of the EPC on the dispersion and lifetime of an electronic or phonon state is determined by the complex electron's or phonon's self-energy, respectively. The real part of the self-energy renormalizes the dispersion, while the imaginary part accounts for the finite lifetime arising from the interactions. One can analyze the renormalized dispersion of an electronic band, and then extract the electron's self-energy and evaluate the lifetime of relevant electrons \cite{McDougall.1995,Reinert.2004}, to determine the possible EPC. Strong EPC generally leads to kinks or quasiparticle peaks in the electronic band near the Fermi energy ($E_F$), as reported by previous angular-resolved photoemission spectroscopy (ARPES) surveys \cite{Hengsberger.1999b,Hengsberger.1999,Rotenberg.2000,Kondo.2013,Yang.2019,Zhou.2008}. Yet, the impacts on electronic properties from EPC usually average the contributions from all involving phonon modes. Consequently, it is rather indirect to assess the contribution from an individual phonon mode from the electron’s perspective only.

EPC also manifests itself from the perspective of phonons, through linewidth broadening \cite{Butler.1979, Wakabayashi.1986}, or the dispersion anomaly which is usually identified as the so-called Kohn anomaly. A Kohn anomaly is a drastic energy softening of a specific phonon branch when its momentum is located at $2k_F$ ($k_F$ the Fermi wave number), resulting from the coupling of the specific phonon mode with electrons near the Fermi surface \cite{Kohn.1959,Kohn.1962}. So Kohn anomaly can vail the quantification of EPC for an individual phonon mode. Kohn anomalies have been observed from phonon dispersion measurements in lots of materials, such as the typical one-dimensional systems with Peierls transitions \cite{Renker.1973,Shirane.1976,Pouget.1991,Guster.2019}, two-dimensional (2D) systems with strong EPC \cite{Politano.2015,Haddad.2018,Weber.2011}, and three-dimensional (3D) bulk systems \cite{Brockhouse.1962,Wong.2003,Agrestini.2004,Baron.2004}.

Similar to the 2D case, the EPC on surfaces, often viewed as quasi-2D, can lead to surface phonon anomalies. For example, the kinks in the phonon dispersions on the (111) surfaces of noble metals \cite{Kaden.1992,Doak.1983,Harten.1985,Harten.1985b,Jayanthi.1987}, are believed to arise from the coupling of surface electronic states to the ion displacement. And the sharp surface phonon anomalies in the hydrogen saturated (110) surfaces of molybdenum and tungsten \cite{Kroger.1997,Hulpke.1993,Kohler.1995, Hulpke.1992,Rotenberg.1998} have the origin due to the interactions with the surface electronic states from the chemical bonding \cite{Benedek.2018b}. More intriguingly, strong Kohn anomalies due to the interactions between surface phonons and the topologically protected surface electrons (or Dirac Fermions) have recently been observed in topological insulators \cite{Zhu.2012,Zhu.2011,Howard.2013} and topological crystalline insulators \cite{Kalish.2019}. In brief, Kohn anomalies should widely exist on a specific surface phonon branch, as long as there exist surface electronic states regardless of their origination.

Recently a new kind of surface states due to the reduction of bulk nonsymmorphic symmetry was discovered in ${\rm ZrSiS}$ \cite{Topp.2017}, which is a candidate of topological nodal line semimetal featured by its nontrivial bulk bands \cite{Schoop.2016,Neupane.2016,Wang.2016,Chen.2017,Hosen.2017,Fu.2019}. ${\rm ZrSiS}$ is considered to be one of the most promising topological materials with potential applications in electronics and spintronics, due to its very large energy range of linear band dispersion \cite{Schoop.2016}, extremely large non-saturating magnetoresistance and high mobility of charge carriers \cite{Ali.2016,Wang.2016,Lv.2016,Singha.2017,Hu.2017,Pezzini.2018}. However, despite the topological nature in its bulk band, it turns out the surface states of ${\rm ZrSiS}$ are not derived from the bulk topology. Instead, they are highly 2D, floating on top of the bulk, merely due to a reduced symmetry at the surface \cite{Topp.2017}. Consequently, it is natural to expect the unique surface states would also interact with phonons and show anomalies in surface phonon dispersions. The coupling between the surface floating 2D electrons and phonons, which could be signified in low dimensional devices, is of exceptional importance for any potential applications, but remains elusive.

Here, employing the technique of high-resolution electron energy loss spectroscopy (HREELS), we systematically mapped out the surface phonon dispersions of ${\rm ZrSiS}$. The results reveal a prominent signature of the surface EPC, manifesting as a strong Kohn anomaly of an optical surface phonon branch with a V-shaped minimum at approximately $2\bm{k}_{\bm{F}}$. Theoretical analysis attributes this strong renormalization of the surface phonon to the interactions with the surface floating 2D electrons of ${\rm ZrSiS}$. Then an average branch-specific EPC constant of 0.15 is obtained from the fitting to the experimental dispersion data.

\section{\label{sec:method}Method}
The ${\rm ZrSiS}$ crystals investigated were grown from the chemical vapor transport method with iodine as the agent. High-purity elemental zirconium $(99.99\%)$, silicon $(99.999\%)$ and sulfur $(99.99\%)$ were firstly mixed together and sintered at 1000 $^{\circ}$C for 5 days to attain polycrystalline powders. Then the polycrystalline powders and iodine were sealed in silica tubes under vacuum in a mass ratio of 1:0.07. The silica tubes were put in a gradient tube furnace with the source powders at 1050 $^{\circ}$C and the cold end at around 950 $^{\circ}$C for 7 days, to acquire rectangular plane-like ${\rm ZrSiS}$ crystals at the cold end.

Single crystalline ${\rm ZrSiS}$ was cleaved {\it in situ} in ultrahigh vacuum (better than $5\times10^{-10}$ Torr) with the surface quality and the crystallographic orientation examined by low energy electron diffraction (LEED). The HREELS measurements were performed in a state-of-the-art spectrometer with the capability of 2D energy-momentum mapping \cite{Zhu.2015}. Due to the special design of our facility, it is able to obtain a phonon spectrum for a certain direction through the Brillouin zone (BZ) in a single measurement without rotating sample, monochromator, or analyzer. All the measurements were done using an incident electron beam with energy of 110 eV and incident angle of 60$^{\circ}$ unless otherwise specified. The energy and momentum resolutions were $\Delta E\sim3$ meV and $\Delta k\sim 0.01\ {\rm \AA}^{-1}$ in this study. The low temperature was reached by continuous liquid helium flow through the manipulator and measured by a silicon diode on the stage behind the sample.

\begin{figure}[t]
\includegraphics[width=0.48\textwidth]{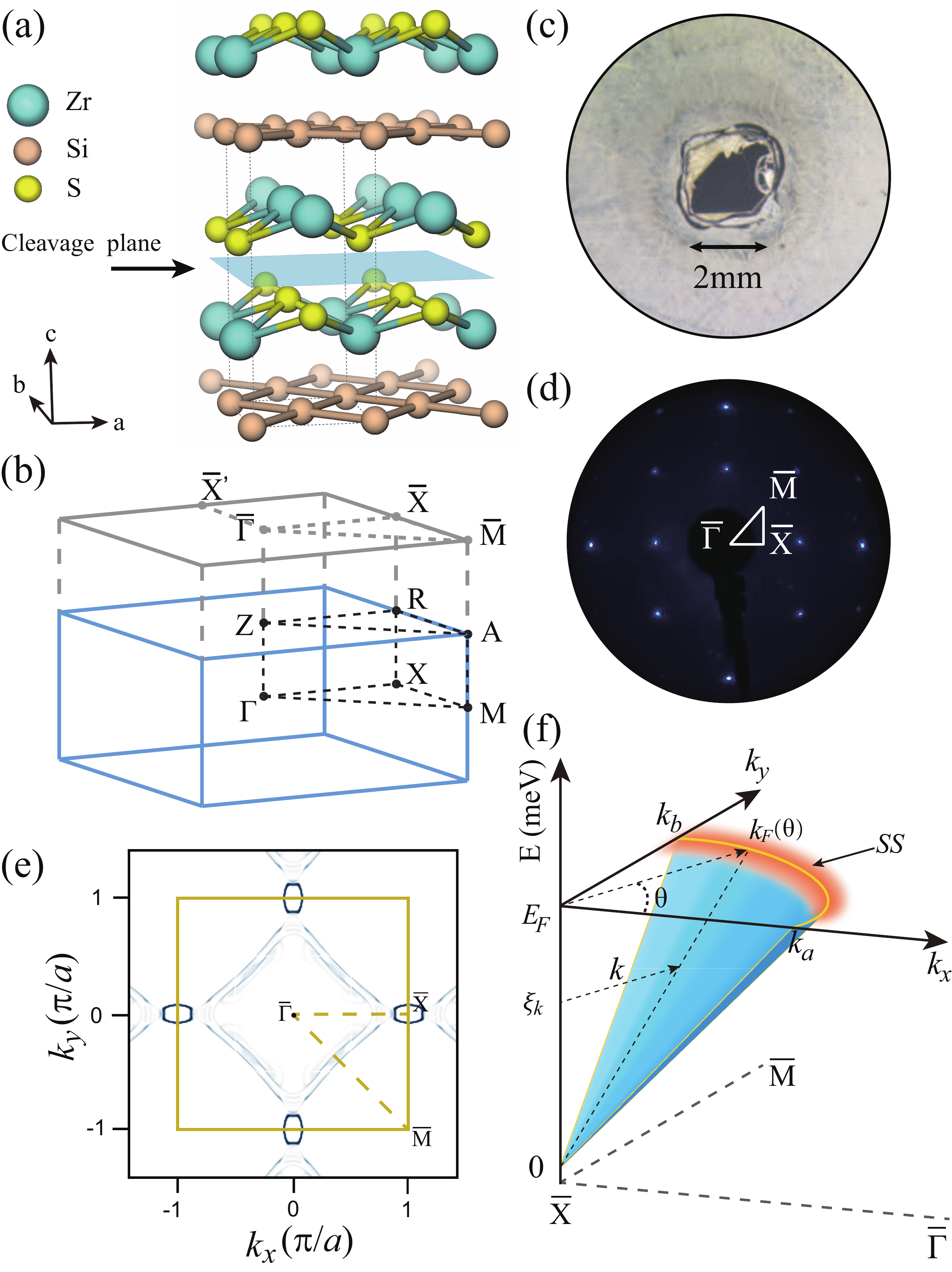}
\caption{\label{fig:structures} \textbf{Sample characteristic and the surface floating 2D states}. (\textbf{a}) Crystal structure of ZrSiS. (\textbf{b}) Illustration of the bulk BZ and the SBZ. (\textbf{c}) The optical image of a sheeny surface of cleaved ZrSiS crystal pasted on a Mo sample holder. (\textbf{d}) Room temperature LEED pattern of a freshly cleaved (001) surface with an incident electron energy of 120 eV. The white triangle shows the least irreducible SBZ. (\textbf{e}) The Fermi surfaces from the surface electronic band calculations. The light blue lines are the projection from the bulk bands, while the dark blue ellipses denote the surface floating 2D states. (\textbf{f}) The cone-like surface floating 2D band around $\bar{\rm X}$ point. The orange arch is the elliptical Fermi surface with major axis $2k_a$ and minor axis $2k_b$.}
\end{figure}

First-principle lattice dynamical calculations were performed to obtain the phonon band structures and the corresponding surface local density of states (LDOS). Here, force constants were calculated by a $(3\times3\times3)$ supercell using Vienna $ab\ initio$ simulation package (VASP) \cite{Kresse.1996}, based on density functional perturbation theory (DFPT) \cite{Baroni.2001}. The generalized gradient approximation (GGA) of Perdew-Burke-Ernzerhof (PBE) \cite{Perdew.1992,Blochl.1994} type was used for the exchange-correlation functions. A plane wave cutoff energy of 420 eV with a $(6\times6\times3)$ Monkhorst-Pack $k$-mesh was employed for the integrations over the BZ. The phonon spectrum was obtained by an open-source package \cite{Togo.2015}, and the LDOS on the (001) surface was calculated by surface Green’s function \cite{Wu.2018}.

\section{\label{sec:res_disc}Results and Discussions}
\subsection{\label{subsec:exp_res}Experimental Results}

${\rm ZrSiS}$ has a tetragonal structure with space group $P4/nmm$ (No.129) and point group $D_{4h}$ \cite{KleinHaneveld.1964,Tremel.1987}, formed by stack of ${\rm S-Zr-Si-Zr-S}$ quintuple layers, with lattice constants $a=b=3.546(2)\ {\rm \AA}$, $c=8.055(4)\ {\rm \AA}$ \cite{Singha.2018}. Two adjacent ${\rm Zr-S}$ layers, with each S atom surrounded by four nearest Zr atoms, are sandwiched by two Si layers extending in the $ab$ plane, as shown in Fig. \ref{fig:structures}(a). Two kinds of nonsymmorphic symmetry can be recognized: a glide mirror in the plane formed by the square nets of the Si atoms and two screw axes $C_{2x}$$(C_{2y})$ along the $a(b)$ directions in the Si layer. The nonsymmorphic symmetry with glide mirror is crucial for the formation of the robust Dirac cone below $E_F$ \cite{Chen.2017} as well as the surface floating 2D bands \cite{Topp.2017}. The weak bonding energy between the two adjacent ${\rm Zr-S}$ layers \cite{Salmankurt.2017} allows easy cleavage along the $ab$ planes to obtain the (001) surface. The breaking of the translational symmetry after cleavage reduces the space group into $P4mm$, corresponding to a four-fold rotational symmetry along the $c$-axis. The bulk Brillouin zone (BZ) is reduced to a square surface Brillouin zone (SBZ), as shown in Fig. \ref{fig:structures}(b). Fig. \ref{fig:structures}(c) and (d) show the optical image and the LEED pattern of a freshly cleaved (001) surface, respectively. Most of the cleaved samples show bright sharp spots with dark background in the LEED patterns, indicating high-quality surfaces obtained for the HREELS measurements. All the LEED patterns show ($1\times1$) square lattices with no signal of surface reconstructions. In Fig. \ref{fig:structures}(e), the calculated Fermi surface on the SBZ illustrates the surface floating 2D bands. Such surface states show as ellipses around $\bar{\rm X}$ at the Fermi energy and shrink nearly linearly to a vertex at around 350 meV below the Fermi level \cite{Topp.2017,Neupane.2016,Hosen.2017}, as demonstrated in Fig. \ref{fig:structures}(f). So we approximate the surface floating 2D states as cone-like band with linear dispersions. The Fermi wave numbers of such cones are $k_a\approx0.15\ {\rm \AA}^{-1}$ along the major axis, and $k_b\approx0.11\ {\rm \AA}^{-1}$ along the minor axis, of the ellipses.

\begin{figure}[t]
\includegraphics[width=0.48\textwidth]{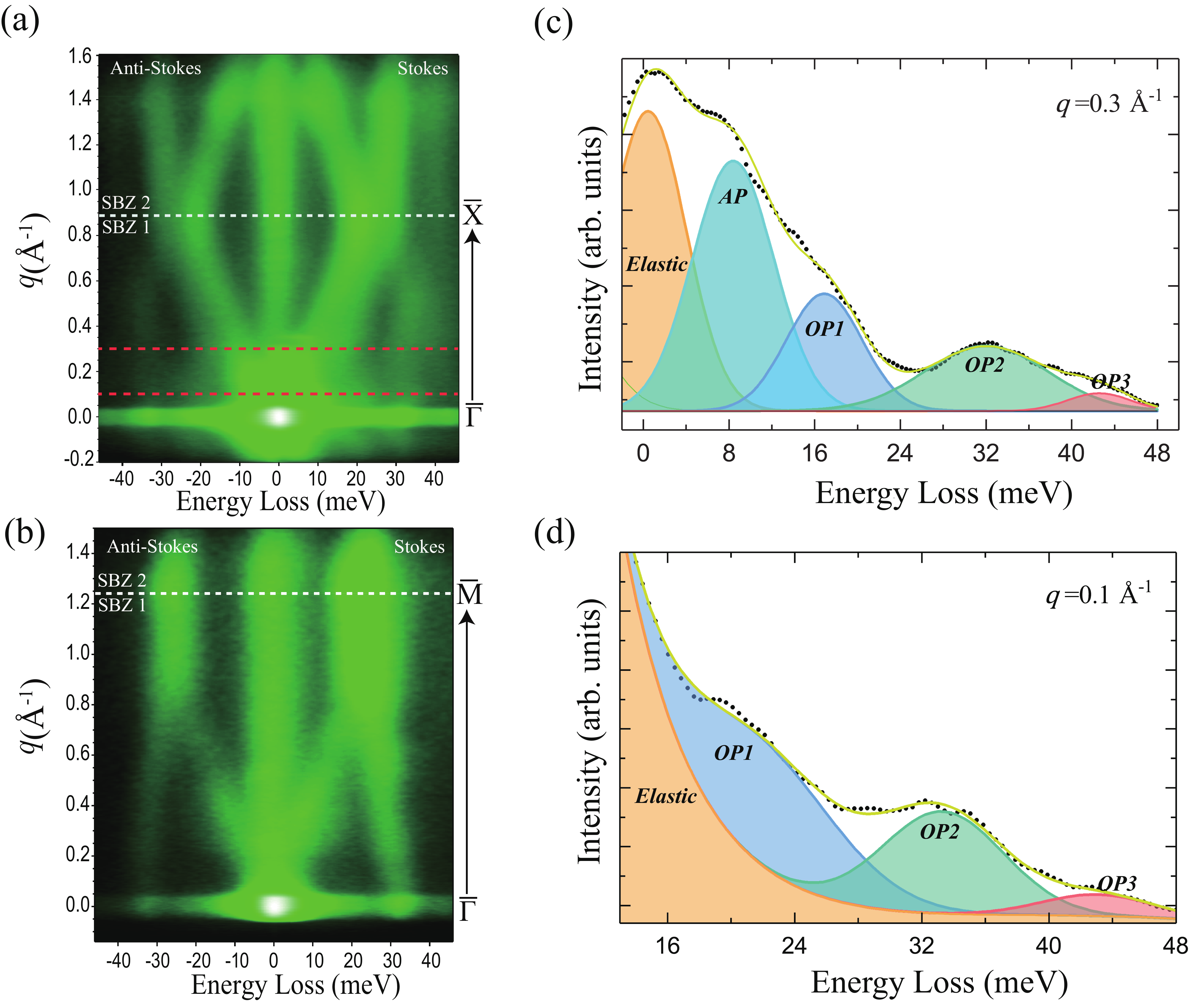}
\caption{\label{fig:hreels_raw} \textbf{HREELS measurement results}. (\textbf{a}), (\textbf{b}) 2D momentum-energy mappings at RT along $\bar{\Gamma}\bar{\rm X}$ and $\bar{\Gamma}\bar{\rm M}$, respectively. The negative energy range corresponds to anti-Stokes peaks (phonon annihilation), while the positive range corresponds to Stokes peaks (phonon creation). The white dashed lines denote the SBZ boundaries. The brightness, representing the signal intensity, is plotted in logarithmical scale. (\textbf{c}), (\textbf{d}) Energy distributed curves along the red dash lines at $q=0.3\ {\rm \AA}^{-1}$ and $q=0.1\ {\rm \AA}^{-1}$ in (a), respectively. The orange background is the elastic peak. Phonon peaks are fitted by Gaussian functions and labelled by AP (acoustic phonon) or OP (optical phonon) according to their dispersions.}
\end{figure}

Figure \ref{fig:hreels_raw}(a) and (b) show the energy and momentum mappings of the surface phonons for ZrSiS obtained from the HREELS measurement at room temperature (RT) along $\bar{\Gamma}\bar{\rm X}$ and $\bar{\Gamma}\bar{\rm M}$, respectively. At least three phonon branches can be discerned in the energy range of 0 - 45 meV. The accurate energy and linewidth information of the phonon modes can be obtained by extracting energy distributed curves (EDCs) from the mapping. For example, two EDCs at $q=0.1\ {\rm \AA}^{-1}$ and $q=0.3\ {\rm \AA}^{-1}$ along $\bar{\Gamma}\bar{\rm X}$ [red dashed lines in Fig. \ref{fig:hreels_raw}(a)], are demonstrated in Fig. \ref{fig:hreels_raw}(c) and (d), respectively. After subtracting the background from elastic scattering, the phonon modes can be distinguished by fitting the energy loss peaks with Gaussian functions. At $q=0.3\ {\rm \AA}^{-1}$, one acoustic phonon (AP) mode and three optical phonon (OP) modes are clearly discerned. At $q=0.1\ {\rm \AA}^{-1}$, only the OP modes are clear, since the AP mode is so close to the elastic peak that its intensity is merged in the strong elastic scattering background.

\subsection{\label{subsec:rev_Kohn}Reveal of the Kohn Anomaly}

\begin{figure*}[t]
\includegraphics[width=0.9\textwidth]{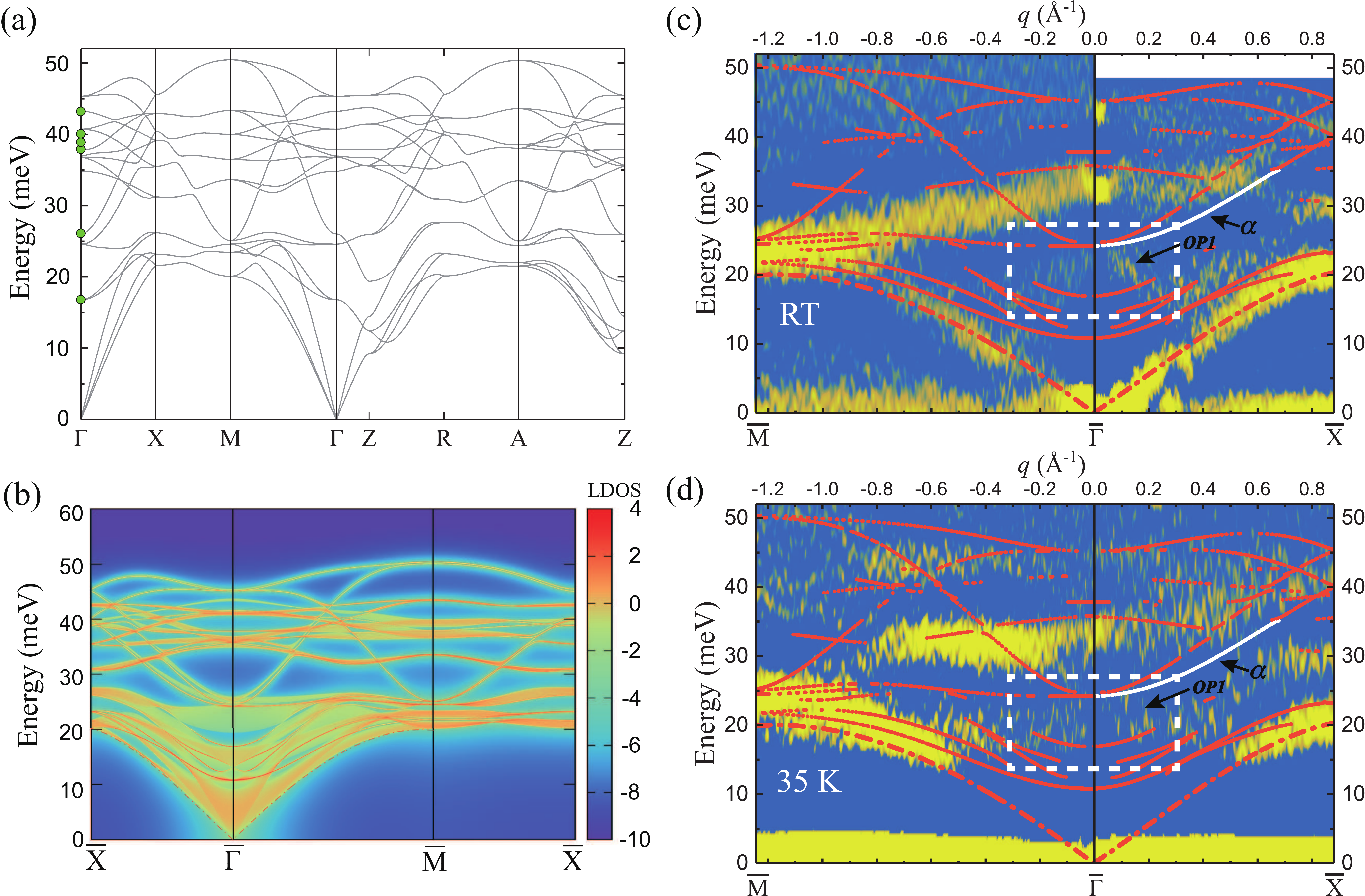}
\caption{\label{fig:phonon_disp} \textbf{Calculated and experimental phonon dispersions}. (\textbf{a}) Calculated bulk phonon dispersions along high symmetry directions of the bulk BZ. The green dots correspond to the Raman-active modes \cite{Singha.2018,Zhou.2017}. (\textbf{b}) Calculated LDOS for phonons of the $(001)$ surface along $\bar{\rm X}-\bar{\Gamma}-\bar{\rm M}-\bar{\rm X}$, with the intensity plotted in logarithmical scale. The yellow shades are the projection of bulk phonon dispersions onto the surface, and red sharp lines represent surface phonons with the LDOS larger than 2.5. The dash-dot lines denote the Rayleigh modes splitting from the bulk acoustic branch. (\textbf{c}) The second derivative image along $\bar{\rm M}-\bar{\Gamma}-\bar{\rm X}$ of the SBZ at RT obtained from the 2D HREELS mapping shown in Fig. \ref{fig:hreels_raw}. The superimposed red lines are the calculated surface phonon dispersions extracted from (b). The white dashed block designates the zone where the anomalous phonon softening occurs. The experimentally observed OP1 mode is compared with the theoretical $\alpha$ mode (white line). (\textbf{d}) The same comparison as (c) with the experimental data obtained at 35 ${\rm K}$.}
\end{figure*}

To analyze the measured phonon dispersions of ${\rm ZrSiS}$ in detail, we carried out first-principle bulk and surface lattice dynamical calculations. Figure \ref{fig:phonon_disp}(a) shows the calculated bulk phonon dispersions. The green dots superimposed on the energy axis denote the locations of the Raman active modes \cite{Singha.2018,Zhou.2017}, indicating good agreement with our calculations. Figure \ref{fig:phonon_disp}(b) shows the dynamical calculation results with the phonon local density of states (LDOS) projected onto the (001) surface using the method of surface Green’s function. The surface phonon modes contribute the highest LDOS manifest as sharp red lines, superimposed on the bulk-projected yellow bands. An elaborate comparison between the calculated and the measured surface phonon dispersions can be carried out by extracting the surface phonon modes from Fig. \ref{fig:phonon_disp}(b) and superimposing them on the second differential images of our experimental spectra. The results are demonstrated in Fig. \ref{fig:phonon_disp}(c) and (d) with experimental data collected at RT and 35 ${\rm K}$, respectively. Although the overall intensity at low temperature is weaker than that at RT, the measured phonon dispersions do not show significant temperature-dependence within the energy resolution, implying the phonon-phonon interactions are negligible. Due to the restriction of selection rules \cite{ibach.1982,Juan.2015}, not all the calculated phonon branches can be measured by HREELS. For those branches that are measured, most of them fit well with the calculations.

The most noticeable feature from the comparison is the optical mode encircled by the white rectangle (OP1 mode) in Fig. \ref{fig:phonon_disp}(c) and (d), which is dramatically discrepant from the calculated surface modes. This peculiar mode OP1 sets out from 25 meV at $\bar{\Gamma}$ point and softens sharply to near 16 meV at $q\approx0.3\ {\rm \AA}^{-1}$ along $\bar{\Gamma}\bar{\rm X}$, while at $q\approx0.25\ {\rm \AA}^{-1}$ along $\bar{\Gamma}\bar{\rm M}$. Then it inclines with increasing $q$ afterward. Although there are several calculated surface modes intersecting with the softened mode OP1 within the white rectangle, most of them are ruled out by the HREELS selection rules \cite{ibach.1982,Juan.2015}. Only the calculated optical surface mode labeled $\alpha$ in Fig. \ref{fig:phonon_disp}(c), coinciding with OP1 in energy at $\bar{\Gamma}$ point, is allowed to be detected by the selection rules. Thus, the observed OP1 mode is designated to the $\alpha$ mode but with energy softened with increasing $q$. At low temperature, the softening of this mode remains to the same extent despite the overall decreased intensity of the entire spectrum. This indicates that the softening originates from EPC rather than phonon-phonon interactions since the latter should be strongly temperature-dependent.

\subsection{\label{subsec:EPC_Model}Electron-phonon Coupling Model}
Similar softening of surface optical phonon modes, interpreted as strong Kohn anomalies, have been observed in quite a few topological materials \cite{Howard.2013,Kalish.2019,Zhu.2012,Zhu.2011}. Such strong Kohn anomalies descend from an abrupt change in the electron screening of atomic vibrations induced by surface electrons, indicating strong surface electron-phonon interactions. For ZrSiS, the surface floating 2D bands around $\bar{\rm X}$ [Fig. \ref{fig:structures}(e)] are the only surface electronic states \cite{Topp.2017,Neupane.2016,Schoop.2016,Wang.2016,Chen.2017,Hosen.2017}. Here, we set up a model in which the surface optical phonon interacts with the cone-like surface floating 2D electrons, to explain the observed Kohn anomaly and describe the detailed EPC picture.

The impact of EPC on phonons can be taken into account using the Dyson equation that describes the relation between perturbed and bare phonons,
\begin{equation}\label{eq:Dyson_equation_def}
\mathcal{D}_s\left(\bm{q},i\omega_n\right)=\frac{\mathcal{D}_s^{\left(0\right)}\left(\bm{q},i\omega_n\right)}{1-\mathcal{D}_s^{\left(0\right)}\left(\bm{q},i\omega_n\right)\Pi\left(\bm{q},\omega_n\right)}
\end{equation}

Here, $\mathcal{D}_s^{(0)}(\bm{q},i\omega_n)$ and $\mathcal{D}_s(\bm{q},i\omega_n)$ are the bare phonon and perturbed phonon Matsubara Green functions, respectively, and $\omega_n$ is the Matsubara frequency for phonons. Under random phase approximation (RPA) the phonon self-energy $\Pi(\bm{q},\omega_n)$ can be substituted by
\begin{equation}\label{eq:SelfEnergy_def}
\Pi\left(\bm{q},\omega_n\right)=\left|g_{\bm{q},s}\right|^2\frac{\mathcal{P}\left(\bm{q},i\omega_n\right)}{\varepsilon\left(\bm{q},i\omega_n\right)}
\end{equation}
where $|g_{\bm{q},s}|$ is the electron-phonon interaction matrix, $\mathcal{P}(\bm{q},i\omega_n)$ is the electron polarization function and $\varepsilon(\bm{q},i\omega_n)=1-v_c(\bm{q})\mathcal{P}(\bm{q},i\omega_n)$ is the RPA dielectric function. For surface electrons,$\ v_c(\bm{q})$ is chosen to be the 2D Fourier transformation of the Coulomb potential, $v_c(\bm{q})=\frac{2\pi e^2}{\kappa|\bm{q}|}$. The electron polarization function can be calculated from the bare electron Matsubara function $\mathcal{G}^{(0)}(\bm{k},ik_n)$ by
\begin{equation}\label{eq:Polarization_Func_def}
\mathcal{P}\left(\bm{q},\omega_n\right)=\frac{1}{\mathcal{A}}\frac{2}{\beta}\sum_{\bm{k}}\sum_{ik_n}{\mathcal{G}^{\left(0\right)}\left(\bm{k},ik_n\right)\mathcal{G}^{\left(0\right)}\left(\bm{k}+\bm{q},ik_n+i\omega_n\right)}
\end{equation}
which describes an electron of momentum $\bm{k}$ and energy $k_n$ scattered to a state of momentum $\bm{k}+\bm{q}$ and energy $k_n+\omega_n$ by a phonon of wave vector $\bm{q}$ and energy $\omega_n$.

\begin{figure}[t]
  \includegraphics[width=0.5\textwidth]{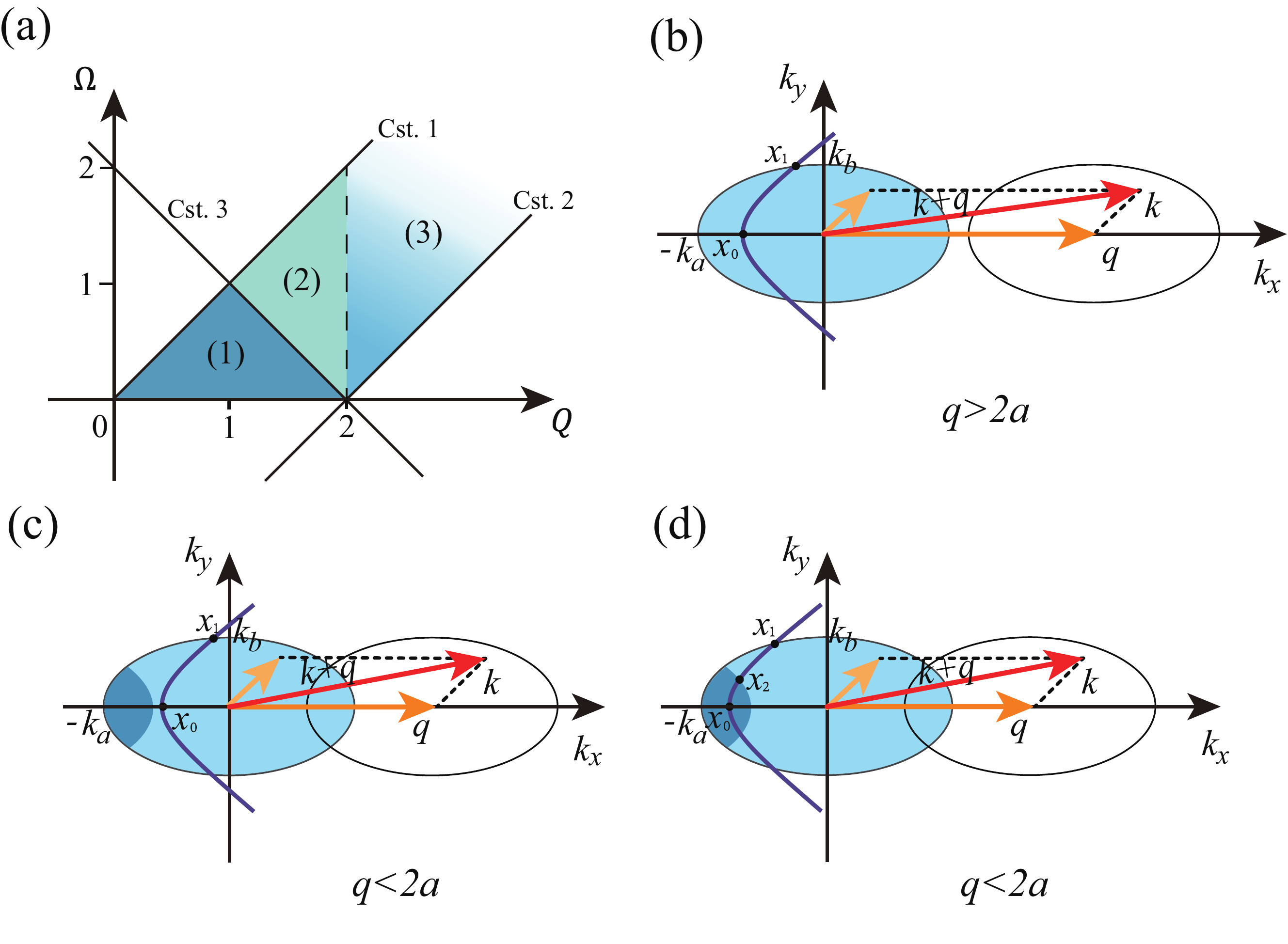}
  \caption{\label{fig:epc_model} \textbf{Illustrations of the electron-phonon scattering process}. (\textbf{a}) The allowed $(\Omega, Q)$ space for phonons defined by the constraint conditions (Cst. 1), (Cst. 2) and (Cst. 3). Regions (3), (2) and (1) correspond to (b), (c) and (d), respectively. (\textbf{b}), (\textbf{c}) and (\textbf{d}) Schematic drawings of the initial and final Fermi ellipses before and after interacting with phonons in three cases of different momentum and energy transfer. The light blue zones denote the allowed area of initial states for electron-phonon interactions, while the dark blue zones are the banned area. The royal blue hyperbolas are the lines that satisfy the conservation of energy.}
\end{figure}

The following analysis focuses on the phonon dispersion along $\bar{\Gamma}\bar{\rm X}$, which involves two orthogonal axes of the Fermi ellipse at two inequivalent $\bar{\rm X}$ [ Fig. \ref{fig:structures}(e)]. Taking phonons with wave vectors along the major axis of the Fermi ellipse for example, due to the constraints of energy and momentum conservation, there are three cases of the electron-phonon scattering processes with different energy and momentum ranges [Fig. \ref{fig:epc_model}(b), (c) and (d)]. The allowed initial states of electrons reside on the royal blue hyperbola within the light blue area. Careful analysis with details for these cases can be found in the Appendix. This analysis yields three constraint conditions for phonons involved in the EPC,
\begin{flushright}
$
\left\{
\begin{aligned}
  \Omega & <Q   \qquad\qquad\qquad & {\rm (Cst.1)}\\
  \Omega & >Q-2 \qquad\qquad\qquad & {\rm (Cst.2)}\\
  \Omega & <2-Q \qquad\qquad\qquad & {\rm (Cst.3)}
\end{aligned}
\right.
$
\end{flushright}
where $\Omega=\frac{\omega}{E_F}$ and $Q=\frac{q}{k_F}$ are the normalized frequency and normalized wave vector of phonons. These constraints require the phonons to fall into the regions (3), (2) and (1) in Fig. \ref{fig:epc_model}(a), corresponding to Fig. \ref{fig:epc_model}(b), (c) and (d), respectively. After an analytical continuation $i\omega_n\rightarrow\ \omega_{\bm{q},s}+i\delta$ in Eqn.(\ref{eq:Polarization_Func_def}), its imaginary part can be obtained by performing the integration within the allowed areas of Fig. \ref{fig:epc_model}(a), and its real part is achieved by conducting a Kramers-Kronig transformation. A similar process can be applied when the phonon wave vector is along the minor axis of Fermi ellipse. The total polarization is just the sum of the two individual polarizations along the two axes.

The renormalized phonon dispersion is given by the real part of the singularity of the Dyson equation in Eqn. (\ref{eq:Dyson_equation_def}),
\begin{equation}\label{eq:phonon_renorm}
    (\hbar\omega_{\bm{q},s})^2=
    (\hbar\omega_{(\bm{q},s)}^{(0)})^2 + 2(\hbar\omega_{\bm{q},s}^{(0)})\left|g_{\bm{q},s}\right|^2 \textup{Re}\left[\frac{\mathcal{P}(\bm{q},\omega_{\bm{q},s})}{\varepsilon(\bm{q},\omega_{\bm{q},s})}\right],
\end{equation}
where $\omega_{\bm{q},s}$ and $\omega_{\bm{q},s}^{(0)}$ correspond to the real and bare frequency for a specific phonon branch $s$ at wave vector $\bm{q}$ with or without EPC, respectively. For the surface phonons, we take the assumption $\left|g_{\bm{q},s}\right|=\sqrt{\frac{N\hbar}{2M\omega_{\bm{q},s}^{\left(0\right)}}}(\gamma_\bot+\frac{\left|\bm{q}\right|}{{2k}_F}\gamma_{||})$ \cite{Zhu.2011} in the momentum range $\bm q<2k_F$, where $M$ is the unit cell mass and $N$ is the number of unit cells. $\gamma_\bot$ and $\gamma_{||}$ are treated as out-of-plane and in-plane interaction parameters to fit Eqn. (\ref{eq:phonon_renorm}) self-consistently. Details of the parameter fitting are described in the Appendix. Here $\omega_{\bm{q},s}^{(0)}$ is obtained from the calculated results of $\alpha$ in Fig. \ref{fig:phonon_disp}(c), and $\omega_{\bm{q},s}$ obtained from the model can be directly compared with the experimentally measured dispersion of the OP1 branch. The renormalized dispersion is depicted in Fig. \ref{fig:epc_parameters}(a).

\subsection{\label{subsec:EPC_Constant}Mode-Specific EPC constant}
From the perspective of phonons, one may assess the branch-specific coupling constant $\lambda_s(\bm{q})$ as a function of momentum $\bm{q}$ for specific phonon branch $s$ by \cite{Butler.1979,Zhu.2012}
\begin{equation}\label{eq:branch_lambda_def}
\lambda_s\left(\bm{q}\right)=-\frac{\textup{Im}\left[\Pi\left(\bm{q},\omega_{\bm{q},s}\right)\right]}{\pi\mathcal{N}\left(E_F\right)\left(\hbar\omega_{\bm{q},s}\right)^2},
\end{equation}
where $\mathcal{N}(E_F)$ is the electron density of states at the Fermi energy.

\begin{figure}[t]
\includegraphics[width=0.48\textwidth]{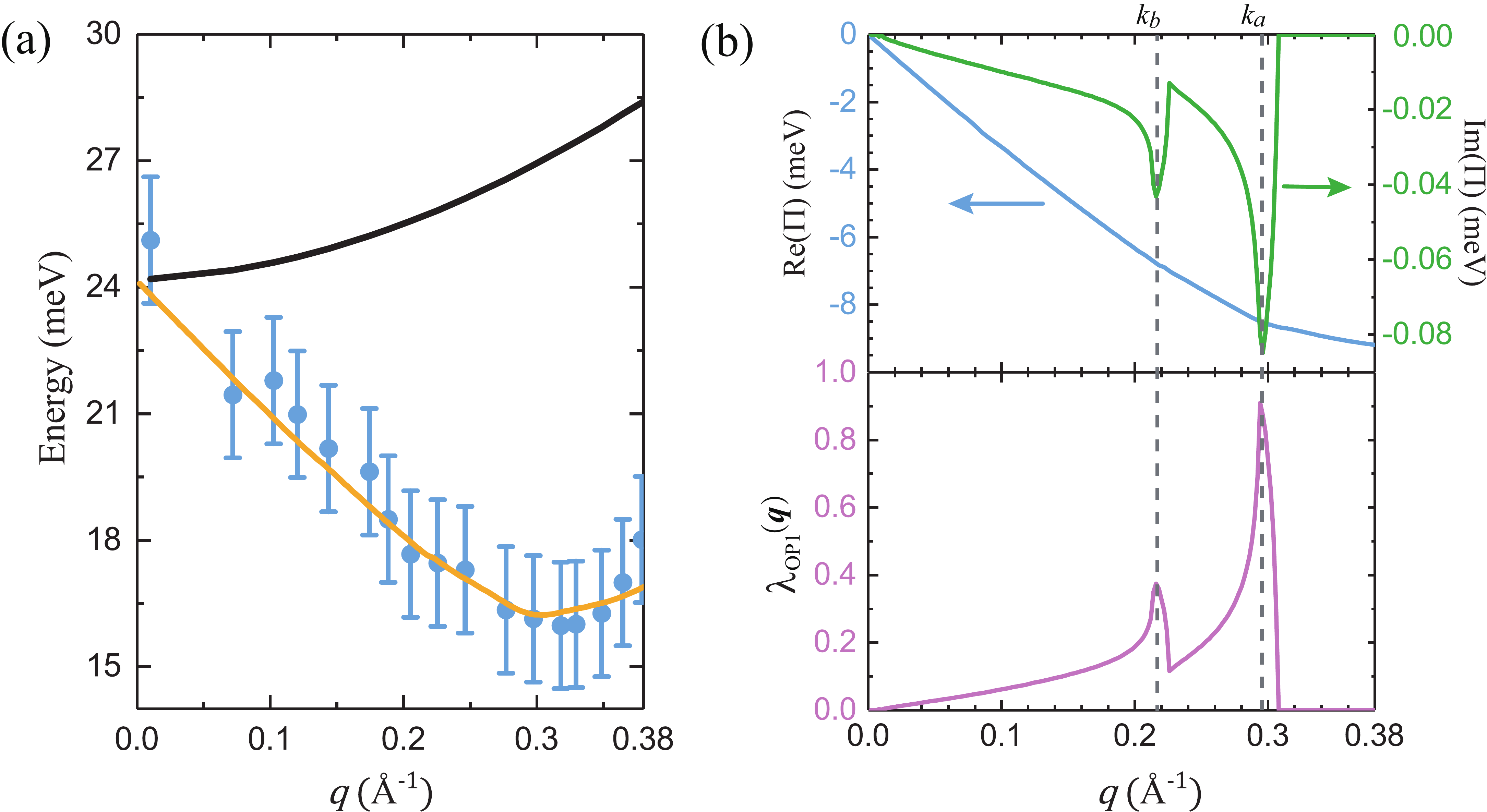}
\caption{\label{fig:epc_parameters} \textbf{Renormalized dispersion and fitting results from the EPC model}. (\textbf{a}) The dispersions of the OP1 mode. The black line is the calculated bare surface phonon dispersion from first principle calculation, while the blue dots with error bars (using the instrument resolution here) are the extracted experimental data. The yellow line is the renormalized dispersion from the EPC model. (\textbf{b}) Calculated real (light blue) and imaginary (green) parts of the phonon self-energy and the EPC constant (magenta) for the OP1 mode.}
\end{figure}

Shown in Fig. \ref{fig:epc_parameters}(b) are the real part and the imaginary part of the phonon self-energy, as well as the mode-specific EPC constant $\lambda_s(\bm{q})$ for the branch OP1 as a function of phonon momentum. The two conspicuous peaks in the imaginary part of the phonon self-energy and $\lambda_{\textup{OP1}}(\bm{q})$ correspond to the two orthogonal axes of the Fermi ellipse. This is different from the conventional 2D Kohn anomaly model \cite{Kohn.1959,Kohn.1962} where an isotropic Fermi circle generates identical $2k_F$ along all the momentum directions. Here, even along one specific direction in the SBZ, $2k_F$ has two maxima at $2k_a$ and $2k_b$. Yet, the physical essence of the EPC is the same, i.e., the EPC constant $\lambda_{\textup{OP1}}(\bm{q})$ declines quickly when $q>2k_a$ or $q>2k_b$ due to the Fermi ellipse, analogous to the case of the Fermi circle with $\lambda(\bm{q})$ declining when $q>2k_F$. The peak at $q\sim2k_a$ is higher than the peak at $q\sim2k_b$, implying a more significant contribution from electrons with momentum along the major axis of the Fermi ellipse. $\lambda_{\textup{OP1}}(\bm{q})$ declines to zero immediately when $q>2k_a$ and the phonon dispersion inclines afterwards.

The averaged EPC constant over ${\bm q}$ for the branch OP1 is evaluated to be $\bar{\lambda}\approx0.15$. This implies the renormalized mass of involved electrons is $m^\ast=(1+\bar{\lambda})m_e\approx1.15m_e$, where $m_e$ is the bare electron mass. The quantum oscillation experiment has confirmed an unconventional mass enhancement in ZrSiS in the range of $1\sim1.5\ m_e$ of electrons within the bulk “dog-bone-like” Fermi pockets under magnetic field \cite{Pezzini.2018}. The enhancement arises presumably from small-momentum density-wave correlations, which requires strongly enhanced Coulomb interactions between electrons \cite{Huh.2016,Roy.2017}. However, our findings here suggest a mass enhancement of electrons of the surface floating band, due to the interaction with surface optical phonons, which cannot be probed by the conductivity quantum oscillations. Different from the bulk electron mass enhancement that was observed only under the magnetic field, the surface EPC discovered here should always exist and will potentially affect any low-dimensional electronic transporting properties. Recently, quasiparticle interference between different faces of surface states around $\bar{\rm X}$ point has been confirmed using a scanning tunneling microscope \cite{Butler.2017,Su.2018,Zhu.2018}, which also supports our proposition that electrons of surface states around $\bar{\rm X}$ interact with optical phonons.

\section{\label{sec:summary}Summary}
In conclusion, we measured the surface phonon dispersions of ${\rm ZrSiS}$, a candidate of topological nodal line semimetal, along the high symmetry directions in the SBZ. Comparing with DFT calculations, we verified an obvious softening of a surface optical mode due to EPC. Detailed theoretical analyses, via an EPC model within RPA, revealed that the strong renormalization of the surface phonon originates from the interactions with the surface floating 2D electrons of ${\rm ZrSiS}$. Under this model, we calculated the polarization function, evaluated the phonon self-energy, and obtained the mode-specific and $q$-dependent EPC constant. The average EPC constant for the softened surface optical phonon branch is $\bar{\lambda}\approx0.15$, indicating the effective mass of renormalized electrons due to surface EPC is $m^\ast\approx1.15\ m_e$. Similar renormalizations of phonons and EPC behaviors are also expected to be observed in isologues like ZrSiSe, ZrSiTe and HfSiSe, for the similarity of structure and consequent surface floating states. The findings and the detailed studies of EPC in this work will be important for potential applications of these semimetal materials.

\section*{\label{sec:acknow}Acknowledgement}
This work was supported by the National Key Research \& Development Program of China (Nos. 2016YFA0302400, 2017YFA0303600, and 2017YFA0302901). X. Z. was supported by the National Natural Science Foundation of China (No. 11874404) and the Youth Innovation Promotion Association of Chinese Academy of Sciences (No. 2016008).
J. G. was partially supported by BAQIS Research Program (No. Y18G09). Y. S. was partially supported by the Beijing Natural Science Foundation (No. Z180008)

\appendix*
\section{\label{appendix}The electron-phonon scattering process}
In this work the interaction between electrons and phonons are analyzed by a standard Green function technology. We now focus on the estimation of Eqn. \ref{eq:Polarization_Func_def}. Using the Matsubara frequency summation relation
$$
\frac{1}{\beta}\sum_{ip_n}{\mathcal{G}^{\left(0\right)}(\bm{p},ip_n)\mathcal{G}^{\left(0\right)}\left(\bm{k},ip_n+i\omega_n\right)}
=\frac{n_F\left(\xi_{\bm{p}}\right)-n_F(\xi_{\bm{k}})}{i\omega_n+\xi_{\bm{p}}-\xi_{\bm{k}}}
$$
one can rewrite the polarization function as
$$
\mathcal{P}\left(\bm{q},\omega_n\right)=\frac{2}{\mathcal{A}}\ \sum_{\bm{k}}\frac{n_F\left(\xi_{\bm{k}}\right)-n_F\left(\xi_{\bm{k}+\bm{q}}\right)}{i\omega_n+\xi_{\bm{k}}-\xi_{\bm{k}+\bm{q}}}
$$
where $n_F(\xi_{\bm{k}})$ is the Fermi distribution of electrons. This formula can be expressed into two parts
\renewcommand\theequation{A.\arabic{equation}}
\begin{equation}
  \begin{split}
    \mathcal{P}\left(\bm{q},\omega_n\right)=
    \frac{1}{\mathcal{A}}\sum_{\bm{k}}\frac{n_F\left(\xi_{\bm{k}}\right)-n_F\left(\xi_{\bm{k}+\bm{q}}\right)}{i\omega_n+\xi_{\bm{k}}-\xi_{\bm{k}+\bm{q}}}\\
    +
    \frac{1}{\mathcal{A}}\sum_{\bm{k}}\frac{n_F\left(\xi_{\bm{k}}\right)-n_F\left(\xi_{\bm{k}+\bm{q}}\right)}{i\omega_n+\xi_{\bm{k}}-\xi_{\bm{k}+\bm{q}}}
  \end{split}
\end{equation}
Make the substitutions $\bm{k}\rightarrow-\bm{k}-\bm{q}$, and $\bm{k}+\bm{q}\rightarrow\bm{-k}$ in the second term and notice that the energy $\xi_{\bm{k}}$ is an even function of $\bm{k}$, thus we have $\xi_{\bm{k}}\rightarrow\xi_{\bm{k}+\bm{q}}$, $\xi_{\bm{k}+\bm{q}}\rightarrow\xi_{\bm{k}}$, $n_F\left(\xi_{\bm{k}}\right)\rightarrow n_F\left(\xi_{\bm{k}+\bm{q}}\right)$, and $n_F\left(\xi_{\bm{k}+\bm{q}}\right)\rightarrow\ n_F\left(\xi_{\bm{k}}\right)$. Then the formula turns into
\begin{equation}
  \begin{split}
    \mathcal{P}\left(\bm{q},\omega_n\right)=&\frac{2}{\mathcal{A}}\sum_{\bm{k}}n_F\left(\xi_{\bm{k}}\right)\left[1-n_F\left(\xi_{\bm{k}+\bm{q}}\right)\right]\cdot\\
                                        &\left(\frac{1}{\xi_{\bm{k}}-\xi_{\bm{k}+\bm{q}}+i\omega_n}-\frac{1}{\xi_{\bm{k}+\bm{q}}-\xi_{\bm{k}}+i\omega_n}\right)
  \end{split}
\end{equation}
Here we choose the Fermi distribution at zero temperature, $n_F\left(\xi_{\bm{k}}\right)=\theta\left(E_F-\xi_{\bm{k}}\right)$. Using the relation $1-\theta\left(E_F-\xi_{\bm{k}+\bm{q}}\right)=\theta(\xi_{\bm{k}+\bm{q}}-E_F)$, and the analytical continuation $i\omega_n\rightarrow\omega+i\delta$, the polarization function turns into
\begin{equation}\label{eq_a:polarization_cal_expand}
  \begin{split}
     \mathcal{P}\left(\bm{q},\omega\right)=&\frac{2}{\mathcal{A}}\sum_{\bm{k}}\theta\left(E_F-\xi_{\bm{k}}\right)\theta\left(\xi_{\bm{k}+\bm{q}}-E_F\right)\cdot\\
                                      &\left(\frac{1}{\omega+\xi_{\bm{k}}-\xi_{\bm{k}+\bm{q}}+i\delta}-\frac{1}{\omega+\xi_{\bm{k}+\bm{q}}-\xi_{\bm{k}}+i\delta}\right)
  \end{split}
\end{equation}
With the relation $\lim \limits_{\delta\rightarrow0}{\frac{1}{x+i\delta}=P\left(\frac{1}{x}\right)-i\pi\delta(x)}$ [here $P()$ means Cauchy Principle Value], we arrive at the imaginary part of the polarization function
\begin{equation}\label{eq_a:im_polarization}
  \begin{split}
    \textup{Im}\left[\mathcal{P}\left(\bm{q},\omega\right)\right]
      =-\frac{2\pi}{\mathcal{A}}\sum_{\bm{k}}\theta\left(E_F-\xi_{\bm{k}}\right)\theta\left(\xi_{\bm{k}+\bm{q}}-E_F\right)\cdot\\
      \left\{\delta\left[\omega-\left(\xi_{\bm{k}+\bm{q}}-\xi_{\bm{k}}\right)\right]-\delta\left[\omega+\left(\xi_{\bm{k}+\bm{q}}-\xi_{\bm{k}}\right)\right]\right\}\\
      =-\frac{1}{2\pi}\int d\bm{k}\theta\left(E_F-\xi_{\bm{k}}\right)\theta\left(\xi_{\bm{k}+\bm{q}}-E_F\right)\cdot\\
      \left\{\delta\left[\omega-\left(\xi_{\bm{k}+\bm{q}}-\xi_{\bm{k}}\right)\right]-\delta\left[\omega+\left(\xi_{\bm{k}+\bm{q}}-\xi_{\bm{k}}\right)\right]\right\}
  \end{split}
\end{equation}

Now, the most important is to obtain the explicit expression of $\xi_{\bm{k}}$ in order to get the value of the integral in Eqn. (\ref{eq_a:im_polarization}). For simplicity, we will take the case with phonon momentum along $\bar{\mathrm{\Gamma}}\bar{\mathrm{X}}$ as an example.

Fermi surface from the surface floating 2D states can be well approximated by ellipses with the expression $k_F(\theta)=\frac{k_ak_b}{\sqrt{k_a^2\sin^2{\theta}+k_b^2\cos^2{\theta}}}$, where $k_a=0.15\ {\rm \AA}^{-1}$ and $k_b=0.11\ {\rm \AA}^{-1}$ are half of the major and half of minor axes of the ellipse, corresponding to the Fermi wave number along $\bar{\Gamma}\bar{\rm X}$ and $\bar{\rm X}\bar{\rm M}$, respectively. And $\theta$ is the angle between the electron momentum and $\bar{\Gamma}\bar{\rm X}$. We choose the vertex of the cone to be energy 0, and thus $E_F=350\ \textup{meV}$. For a given $k$ along the direction with angle $\theta$ [Fig. \ref{fig:structures}(f)], we get the expression of the electron dispersion
\begin{equation}
  \begin{split}
     \xi_{\bm{k}} & =\frac{E_F}{k_F\left(\theta\right)}\left|\bm{k}\right| \\
                & =\frac{E_F}{k_ak_b}\sqrt{k_a^2\sin^2{\theta}+k_b^2\cos^2{\theta}}\left|\bm{k}\right|
  \end{split}
\end{equation}

After interacting with a phonon mode with momentum $\bm{q}$ along $\bar{\Gamma}\bar{\rm X}$, the electrons transfer from $\bm{k}$ (initial states) into $\bm{k}+\bm{q}$ (final states) with angle $\theta^\prime$ (Fig. \ref{fig_a:ep_scattering}). A simple geometry derivation yields
\begin{equation}
  \begin{split}
     \xi_{\bm{k}+\bm{q}} & =\frac{E_F}{k_F\left(\theta^\prime\right)}\left|\bm{k}+\bm{q}\right| \\
                         & =\frac{E_F}{k_ak_b}\sqrt{k_a^2\left(k\sin{\theta}\right)^2+k_b^2\left(k\cos{\theta}+q\right)^2}
  \end{split}
\end{equation}

\setcounter{figure}{0}
\renewcommand\thefigure{A\arabic{figure}}
\begin{figure}[t]
\includegraphics[width=0.3\textwidth]{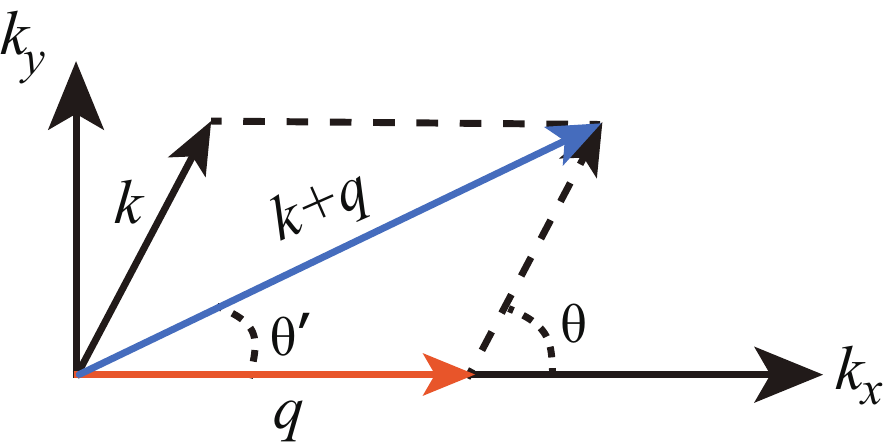}
\caption{\label{fig_a:ep_scattering} \textbf{Schematics of the scattering momentum}. A simple illustration of the relation between momentums of electrons before $(\bm{k})$ and after $(\bm{k}+\bm{q})$ interacting with phonons $\bm{q}$.}
\end{figure}

In orthogonal coordinates, they are expressed as
\begin{equation}
  \begin{split}
     \xi_{\bm{k}} & =E_F\sqrt{\left(\frac{k_x}{k_a}\right)^2+\left(\frac{k_y}{k_b}\right)^2} \\
     \xi_{\bm{k}+\bm{q}} & =E_F\sqrt{\left(\frac{k_x+q}{k_a}\right)^2+\left(\frac{k_y}{k_b}\right)^2}
  \end{split}
\end{equation}

Let's return to Eqn. (\ref{eq_a:im_polarization}). For
$\xi_{\bm{k}}$ has a linear dispersion, the step functions
$\theta(E_F-\xi_{\bm{k}})$ and $\theta(\xi_{\bm{k}+\bm{q}}-E_F)$
require $\left|\bm{k}\right|<k_F$ and
$\left|\bm{k}+\bm{q}\right|>k_F$. There are two situations:
\begin{itemize}
  \item $\left|\bm{q}\right|>2k_F$, all electrons inside the Fermi ellipse can be the initial states satisfying the requirements [Fig. \ref{fig:epc_model}(b)].
  \item $\left|\bm{q}\right|<2k_F$, only electrons outside the banned region of the Fermi ellipse can be the initial states satisfying the requirements [light blue region of Fig. \ref{fig:epc_model}(c) and (d)]. The banned region is the intersection area of initial and final ellipses, shifted by $-q$.
\end{itemize}

There are two $\delta$ functions in Eqn. (\ref{eq_a:im_polarization}), $\delta[\omega-(\xi_{\bm{k}+\bm{q}}-\xi_{\bm{k}})]$ and $\delta[\omega+(\xi_{\bm{k}+\bm{q}}-\xi_{\bm{k}})]$. Only one of these two functions can be nonzero with a given $\omega$, and it is reasonable to choose the former one since it supports positive phonon energy. In fact, the $\delta$ function is the conservation of energy between the initial and final states. Only those electrons with $\bm{k}$ that satisfy
\begin{equation}\label{eq_a:hyperbola}
  \omega-\left(\xi_{\bm{k}+\bm{q}}-\xi_{\bm{k}}\right)=0
\end{equation}
contribute to the integral in Eqn. (\ref{eq_a:im_polarization}), which defines a hyperbolic curve in the momentum plane $k_xk_y$ [royal blue curves in Fig. \ref{fig:epc_model}(b), (c) and (d)].

Considering the mirror symmetry with respect to $k_x$, the integral in Eqn. (\ref{eq_a:im_polarization}) can be rewritten as
\begin{equation}
  \begin{split}
     &\textup{Im}[\mathcal{P}\left(\bm{q},\omega\right)] = -\frac{1}{\pi}\int_{k_x,k_y \in R} dk_xdk_y
     \delta\{\omega-\\
     &E_F[\sqrt{\left(\frac{x+q}{k_a}\right)^2+\left(\frac{y}{k_b}\right)^2}-\sqrt{\left(\frac{x}{k_a}\right)^2+\left(\frac{y}{k_b}\right)^2}]\}
  \end{split}
\end{equation}
where the zone $R$ is the allowed scattering zone in the momentum space above $k_x$ axis [light blue part above $k_x$ in Fig. \ref{fig:epc_model}(b), (c) and (d)]. A simple projection with
$$
\left\{
  \begin{aligned}
         m &= \frac{k_x}{k_a},      &n &= \frac{k_y}{k_b} \\
    \Omega &= \frac{\omega}{E_F}, &Q &= \frac{q}{k_a}
  \end{aligned}
\right.
$$
yields
\begin{equation}\label{eq_a:im_polarization_project}
  \begin{split}
     \textup{Im}\left[\mathcal{P}\left(\bm{q},\omega\right)\right] & =-\frac{k_ak_b}{\pi E_F}\iint_{m,n\in R^\prime}dmdn\cdot\\ &\delta\left\{\Omega-\left[\sqrt{\left(m+Q\right)^2+n^2}-\sqrt{m^2+n^2}\right]\right\}
  \end{split}
\end{equation}
where $R^\prime$ is the mapping area of $R$. And the hyperbolic curve function (\ref{eq_a:hyperbola}) turns to
\begin{equation}\label{eq_a:hyperbola_project}
  \Omega-\left[\sqrt{\left(m+Q\right)^2+n^2}-\sqrt{m^2+n^2}\right]=0
\end{equation}

The Fermi surface turns to a circle
\begin{equation}
  m^2+n^2=1
\end{equation}

Now, we are going to derive the constraint condition for $\Omega$ and $Q$. The hyperbola described by Eqn. (\ref{eq_a:hyperbola_project}) intersects with the horizontal axis at $m_0=\frac{\Omega-Q}{2}$, with the Fermi circle at $m_1=\frac{2\Omega+{\Omega}^2}{2Q}-\frac{Q}{2}$ and with the banned region at $m_2=\frac{2\Omega-{\Omega}^2}{2Q}-\frac{Q}{2}$ (In Fig. \ref{fig:epc_model}(b), (c) and (d), $x_0 = m_0 k_a$, $x_1 = m_1 k_a$, $x_2 = m_2 k_a$ are the horizontal values of the intersection points). If the frequency of phonons ($\omega$ or $\Omega$) is too low or too high, the hyperbola will not intersect with the Fermi circle or just does not exist. To make sure the hyperbola exists, its real axis length should be shorter than the focus length
\setcounter{equation}{0}
\renewcommand\theequation{Cst.\arabic{equation}}
\begin{equation}\label{eq_a:osq}
    \Omega<Q
\end{equation}
This is the first constraint condition for $\Omega$ and $Q$. To make sure the hyperbola intersects with the Fermi circle, we have $m_0>-1$, i.e.
\begin{equation}\label{eq_a:obqm2}
  \Omega>Q-2
\end{equation}

Because the hyperbola is right-oriented, we must have $m_1>m_0$ or $m_2>m_0$ [Fig. \ref{fig:epc_model} (d)]. These two conditions also yield the constraint condition \ref{eq_a:obqm2}. For the case where $q<2k_a\ (Q<2)$, this condition is always satisfied because $\Omega$ is positive. But for the case where $q>2k_a\ (Q>2)$, the situation is different. This means phonon modes with small energy will not participate in the electron-phonon interaction with momentum transfer bigger than $2k_a$.

For $q<2k_a\ \left(Q<2\right)$, the hyperbola may enter the banned zone [Fig. \ref{fig:epc_model}(d)]. In this case we have $m_0<1-Q$, i.e
\begin{equation}\label{eq_a:os2mq}
  \Omega<2-Q
\end{equation}

Constraint conditions (\ref{eq_a:osq}), (\ref{eq_a:obqm2}) and (\ref{eq_a:os2mq}) help to define the allowed zone for phonon energy and momentum transfer. The permitted value area of $\Omega$ and $Q$ is illustrated in Fig. \ref{fig:epc_model}(a). The region (3), (2) and (1) correspond to Fig. \ref{fig:epc_model}(b), (c) and (d), respectively.

After some algebra with $n$ integrated out in (\ref{eq_a:im_polarization_project}), we get
\setcounter{equation}{12}
\renewcommand\theequation{A.\arabic{equation}}

\begin{equation}
  \begin{split}
     \textup{Im}\left[\mathcal{P}\left(\bm{q},\omega\right)\right]=-\frac{k_ak_b}{\pi E_F}\int_{m\in R^\prime}{dm\frac{1}{\Omega^2}\cdot}\\
     \frac{Q^2\left(m+\frac{Q}{2}\right)^2-\frac{\Omega^4}{4}}{\sqrt{Q^2-\Omega^2}\sqrt{\left(m+\frac{Q}{2}\right)^2-\frac{\Omega^2}{4}}}
  \end{split}
\end{equation}
With substitution $t=m+Q/2$, we arrive at
\begin{equation}
  \textup{Im}\left[\mathcal{P}\left(\bm{q},\omega\right)\right]=F\left(\bm{q},\omega,t_{ul}\right)-F(\bm{q},\omega,t_{ll})
\end{equation}
where
\begin{equation}
  \begin{split}
    F\left(\bm{q},\omega,t\right)=&-\frac{k_ak_b}{\pi E_F}\left[\frac{Q^2}{2\Omega^2\sqrt{Q^2-\Omega^2}}t\sqrt{t^2-\frac{\Omega^2}{4}}\right.\\
                                  &\left. + \frac{Q^2-2\Omega^2}{8\sqrt{Q^2-\Omega^2}}\ln{\left(t+\sqrt{t^2-\frac{\Omega^2}{4}}\right)}\right]
  \end{split}
\end{equation}
and $t_{ul}$ and $t_{ll}$ are the respective upper and lower integral limits.

We notice that in different regions of Fig. \ref{fig:epc_model}(a) the integral variables of Eqn. (\ref{eq_a:im_polarization_project}) run across different range. The upper limit of all three regions is $t_1=m_1+\frac{Q}{2}=\frac{2\Omega+\Omega^2}{2Q}$. The lower limit for region (2) and (3) is $t_0=m_0+\frac{Q}{2}=\frac{\Omega}{2}$,  while for region (1) is $t_2=m_2+\frac{Q}{2}=\frac{2\Omega-\Omega^2}{2Q}$. Considering the constraint conditions, the integral in the three regions can be expressed as
\begin{equation}
  \begin{split}
     I_1(\bm{q},\omega) = & \theta\left(Q-\Omega\right)\theta\left(2-Q\right)\theta\left[2-\left(\Omega+Q\right)\right]\cdot \\
                          & [F\left(\bm{q},\omega,t_1\right)-F\left(\bm{q},\omega,t_2\right)]\\
     I_2(\bm{q},\omega) = & \theta\left(Q-\Omega\right)\theta\left(2-Q\right)\theta\left[\left(\Omega+Q\right)-2\right]\cdot \\
                          & [F\left(\bm{q},\omega,t_1\right)-F\left(\bm{q},\omega,t_0\right)]\\
     I_3(\bm{q},\omega) = & \theta\left(Q-\Omega\right)\theta\left(Q-2\right)\theta\left[2-\left(Q-\Omega\right)\right]\cdot\\
                          & [F\left(\bm{q},\omega,t_1\right)-F\left(\bm{q},\omega,t_0\right)]
  \end{split}
\end{equation}
respectively. And the total imaginary part of $\mathcal{P}\left(\bm{q},\omega\right)$ is $\textup{Im}\left[\mathcal{P}\left(\bm{q},\omega\right)\right]=I_1(\bm{q},\omega)+I_2(\bm{q},\omega)+I_3(\bm{q},\omega)$. The experimentally corresponding polarization function is $\textup{Im}\left[\mathcal{P}\left(\bm{q},\omega_{\bm{q},s}\right)\right]=I_1(\bm{q},\omega_{\bm{q},s})+I_2(\bm{q},\omega_{\bm{q},s})+I_3(\bm{q},\omega_{\bm{q},s})$, where $\hbar\omega_{\bm{q},s}$ is the experimental phonon energy.

With the help of Kramers-Kronig transformation, the real part of the polarization function at $\hbar\omega_{\bm{q},s}$ can be calculated
\begin{equation}
  \textup{Re}\left[\mathcal{P}\left(\bm{q},\omega_{\bm{q},s}\right)\right]=\frac{2}{\pi}\int_{0}^{\infty}\frac{\omega \textup{Im}[\mathcal{P}(\bm{q},\omega)]}{\omega^2-{\omega_{\bm{q},s}}^2}d\omega
\end{equation}

Then we can calculate both the imaginary and real parts of the RPA dielectric function by $\varepsilon(\bm{q},i\omega_n)=1-\frac{2\pi e^2}{\kappa|\bm{q}|}\mathcal{P}(\bm{q},i\omega_n)$.

In the small phonon momentum range $(q<2k_F)$, we can assume the following expression of electron-phonon interaction matrix $\left|g_{\bm{q},s}\right|$ for an optical phonon mode \cite{Zhu.2011}
$$
\left|g_{\bm{q},s}\right|=\sqrt{\frac{N\hbar}{2M\omega_{\bm{q},s}^{\left(0\right)}}}(\gamma_\bot+\frac{\left|\bm{q}\right|}{{2k}_F}\gamma_{||})
$$
where $M$ is the unit cell mass and $N$ is the number of unit cells in a sample, $\gamma_{||}$ is the interaction constant within $ab$ plane and $\gamma_\bot$ is the constant out of plane. Thus Eqn. (\ref{eq:phonon_renorm}) turns to
\begin{equation}
\left(\hbar\omega_{\bm{q},s}\right)^2=\left(\hbar\omega_{\bm{q},s}^{\left(0\right)}\right)^2+
\frac{N\hbar^2}{M}\left(\gamma_\bot\right)^2\left(1+\frac{\left|\bm{q}\right|}{{2k}_F}\frac{\gamma_{||}}{\gamma_\bot}\right)^2\textup{Re}\left[\frac{\mathcal{P}\left(\bm{q},\omega_{\bm{q},s}\right)}{\varepsilon\left(\bm{q},\omega_{\bm{q},s}\right)}\right]
\end{equation}
where we leave $\frac{N\hbar^2}{M}\left(\gamma_\bot\right)^2$ and $\frac{\gamma_{||}}{\gamma_\bot}$ as fitting parameters.

The best fitting parameters with the experimental data are
$$
\frac{N\hbar^2}{M}\left(\gamma_\bot\right)^2={1.51\times10}^8\ \left(\textup{meV}\right)^3\cdot{\rm \AA}^2
$$
and $$
\frac{\gamma_{||}}{\gamma_\bot}=0.025.
$$
The fitted renormalized dispersion is illustrated as yellow line in Fig. \ref{fig:epc_parameters}(a).

With the fitting parameters, we can evaluate the real and imaginary parts of the phonon self-energy by

\begin{equation}
  \begin{split}
    \textup{Re}[\Pi\left(\bm{q},\omega_{\bm{q},\bm{s}}\right)]
    &=\left|g_{\bm{q},s}\right|^2
      \textup{Re}[\frac{\mathcal{P}\left(\bm{q},\omega_{\bm{q},\bm{s}}\right)}{\varepsilon\left(\bm{q},\omega_{\bm{q},\bm{s}}\right)}]\\
    &=\frac{N\hbar}{2M\omega_{\bm{q},s}^{\left(0\right)}}
      \gamma_\bot^2(1+\frac{|\bm{q}|}{2k_F}\frac{\gamma_{||}}{\gamma_\bot})^2
      \textup{Re}[\frac{\mathcal{P}(\bm{q},\omega_{\bm{q},s})}{\varepsilon(\bm{q},\omega_{\bm{q},s})}]\\
    \textup{Im}[\Pi\left(\bm{q},\omega_{\bm{q},\bm{s}}\right)]
    &=\frac{N\hbar}{2M\omega_{\bm{q},s}^{\left(0\right)}}
      \gamma_\bot^2(1+\frac{|\bm{q}|}{2k_F}\frac{\gamma_{||}}{\gamma_\bot})^2
      \textup{Im}[\frac{\mathcal{P}(\bm{q},\omega_{\bm{q},s})}{\varepsilon(\bm{q},\omega_{\bm{q},s})}]
  \end{split}
\end{equation}

Finally, the EPC constant can be obtained from Eqn. (\ref{eq:branch_lambda_def}), where we take $\mathcal{N}\left(E_F\right)=0.1171\ /\textup{eV}$ for ZrSiS from the DFT calculations. The results are plotted in Fig. \ref{fig:epc_parameters}(b).

\bibliography{EPC_ZrSiS}

\begin{thebibliography}{73}%
\makeatletter
\providecommand \@ifxundefined [1]{%
 \@ifx{#1\undefined}
}%
\providecommand \@ifnum [1]{%
 \ifnum #1\expandafter \@firstoftwo
 \else \expandafter \@secondoftwo
 \fi
}%
\providecommand \@ifx [1]{%
 \ifx #1\expandafter \@firstoftwo
 \else \expandafter \@secondoftwo
 \fi
}%
\providecommand \natexlab [1]{#1}%
\providecommand \enquote  [1]{``#1''}%
\providecommand \bibnamefont  [1]{#1}%
\providecommand \bibfnamefont [1]{#1}%
\providecommand \citenamefont [1]{#1}%
\providecommand \href@noop [0]{\@secondoftwo}%
\providecommand \href [0]{\begingroup \@sanitize@url \@href}%
\providecommand \@href[1]{\@@startlink{#1}\@@href}%
\providecommand \@@href[1]{\endgroup#1\@@endlink}%
\providecommand \@sanitize@url [0]{\catcode `\\12\catcode `\$12\catcode
  `\&12\catcode `\#12\catcode `\^12\catcode `\_12\catcode `\%12\relax}%
\providecommand \@@startlink[1]{}%
\providecommand \@@endlink[0]{}%
\providecommand \url  [0]{\begingroup\@sanitize@url \@url }%
\providecommand \@url [1]{\endgroup\@href {#1}{\urlprefix }}%
\providecommand \urlprefix  [0]{URL }%
\providecommand \Eprint [0]{\href }%
\providecommand \doibase [0]{https://doi.org/}%
\providecommand \selectlanguage [0]{\@gobble}%
\providecommand \bibinfo  [0]{\@secondoftwo}%
\providecommand \bibfield  [0]{\@secondoftwo}%
\providecommand \translation [1]{[#1]}%
\providecommand \BibitemOpen [0]{}%
\providecommand \bibitemStop [0]{}%
\providecommand \bibitemNoStop [0]{.\EOS\space}%
\providecommand \EOS [0]{\spacefactor3000\relax}%
\providecommand \BibitemShut  [1]{\csname bibitem#1\endcsname}%
\let\auto@bib@innerbib\@empty
\bibitem [{\citenamefont {Grimvall}(1981)}]{Grimvall.1981}%
  \BibitemOpen
  \bibfield  {author} {\bibinfo {author} {\bibfnamefont {G.}~\bibnamefont
  {Grimvall}},\ }\href@noop {} {\emph {\bibinfo {title} {The Electron-Phonon
  Interaction in Metals}}},\ Vol.~\bibinfo {volume} {8}\ (\bibinfo  {publisher}
  {North-Holland Amsterdam},\ \bibinfo {year} {1981})\BibitemShut {NoStop}%
\bibitem [{\citenamefont {Maksimov}\ \emph {et~al.}(1997)\citenamefont
  {Maksimov}, \citenamefont {Savrasov},\ and\ \citenamefont
  {Savrasov}}]{Maksimov.1997}%
  \BibitemOpen
  \bibfield  {author} {\bibinfo {author} {\bibfnamefont {E.~G.}\ \bibnamefont
  {Maksimov}}, \bibinfo {author} {\bibfnamefont {D.~Y.}\ \bibnamefont
  {Savrasov}},\ and\ \bibinfo {author} {\bibfnamefont {S.~Y.}\ \bibnamefont
  {Savrasov}},\ }\bibfield  {title} {\bibinfo {title} {\textit{The
  Electron-phonon Interaction and the Physical Properties of Metals}},\ }\href
  {https://doi.org/10.1070/pu1997v040n04abeh000226} {\bibfield  {journal}
  {\bibinfo  {journal} {Phys. Usp.}\ }\textbf {\bibinfo {volume} {40}},\
  \bibinfo {pages} {337} (\bibinfo {year} {1997})}\BibitemShut {NoStop}%
\bibitem [{\citenamefont {Pavarini}\ \emph {et~al.}(2017)\citenamefont
  {Pavarini}, \citenamefont {Koch}, \citenamefont {Scalettar},\ and\
  \citenamefont {Martin}}]{Pavarini.2017}%
  \BibitemOpen
  \bibinfo {editor} {\bibfnamefont {E.}~\bibnamefont {Pavarini}}, \bibinfo
  {editor} {\bibfnamefont {E.}~\bibnamefont {Koch}}, \bibinfo {editor}
  {\bibfnamefont {R.}~\bibnamefont {Scalettar}},\ and\ \bibinfo {editor}
  {\bibfnamefont {R.}~\bibnamefont {Martin}},\ eds.,\ \href
  {http://juser.fz-juelich.de/record/837488} {\emph {\bibinfo {title} {{T}he
  {P}hysics of {C}orrelated {I}nsulators, {M}etals, and {S}uperconductors}}},\
  \bibinfo {series} {Schriften des Forschungszentrums Jülich Reihe Modeling
  and Simulation}, Vol.~\bibinfo {volume} {7}\ (\bibinfo  {publisher}
  {Forschungszentrum Jülich GmbH Zentralbibliothek, Verlag},\ \bibinfo
  {address} {Jülich},\ \bibinfo {year} {2017})\ p.\ \bibinfo {pages}
  {450}\BibitemShut {NoStop}%
\bibitem [{\citenamefont {Giustino}(2017)}]{Giustino.2017}%
  \BibitemOpen
  \bibfield  {author} {\bibinfo {author} {\bibfnamefont {F.}~\bibnamefont
  {Giustino}},\ }\bibfield  {title} {\bibinfo {title} {\textit{Electron-phonon
  Interactions from First Principles}},\ }\href
  {https://doi.org/10.1103/RevModPhys.89.015003} {\bibfield  {journal}
  {\bibinfo  {journal} {Rev. Mod. Phys.}\ }\textbf {\bibinfo {volume} {89}},\
  \bibinfo {pages} {015003} (\bibinfo {year} {2017})}\BibitemShut {NoStop}%
\bibitem [{\citenamefont {McDougall}\ \emph {et~al.}(1995)\citenamefont
  {McDougall}, \citenamefont {Balasubramanian},\ and\ \citenamefont
  {Jensen}}]{McDougall.1995}%
  \BibitemOpen
  \bibfield  {author} {\bibinfo {author} {\bibfnamefont {B.~A.}\ \bibnamefont
  {McDougall}}, \bibinfo {author} {\bibfnamefont {T.}~\bibnamefont
  {Balasubramanian}},\ and\ \bibinfo {author} {\bibfnamefont {E.}~\bibnamefont
  {Jensen}},\ }\bibfield  {title} {\bibinfo {title} {\textit{Phonon
  Contribution to Quasiparticle Lifetimes in {Cu} Measured by Angle-resolved
  Photoemission}},\ }\href {https://doi.org/10.1103/PhysRevB.51.13891}
  {\bibfield  {journal} {\bibinfo  {journal} {Phys. Rev. B}\ }\textbf {\bibinfo
  {volume} {51}},\ \bibinfo {pages} {13891} (\bibinfo {year}
  {1995})}\BibitemShut {NoStop}%
\bibitem [{\citenamefont {Reinert}\ \emph {et~al.}(2004)\citenamefont
  {Reinert}, \citenamefont {Eltner}, \citenamefont {Nicolay}, \citenamefont
  {Forster}, \citenamefont {Schmidt},\ and\ \citenamefont
  {H{\"u}fner}}]{Reinert.2004}%
  \BibitemOpen
  \bibfield  {author} {\bibinfo {author} {\bibfnamefont {F.}~\bibnamefont
  {Reinert}}, \bibinfo {author} {\bibfnamefont {B.}~\bibnamefont {Eltner}},
  \bibinfo {author} {\bibfnamefont {G.}~\bibnamefont {Nicolay}}, \bibinfo
  {author} {\bibfnamefont {F.}~\bibnamefont {Forster}}, \bibinfo {author}
  {\bibfnamefont {S.}~\bibnamefont {Schmidt}},\ and\ \bibinfo {author}
  {\bibfnamefont {S.}~\bibnamefont {H{\"u}fner}},\ }\bibfield  {title}
  {\bibinfo {title} {\textit{The Electron-phonon Self-energy of Metallic
  Systems Determined by Angular Resolved High-resolution Photoemission}},\
  }\href {https://doi.org/10.1016/j.physb.2004.06.013} {\bibfield  {journal}
  {\bibinfo  {journal} {Physica B}\ }\textbf {\bibinfo {volume} {351}},\
  \bibinfo {pages} {229} (\bibinfo {year} {2004})}\BibitemShut {NoStop}%
\bibitem [{\citenamefont {Hengsberger}\ \emph
  {et~al.}(1999{\natexlab{a}})\citenamefont {Hengsberger}, \citenamefont
  {Purdie}, \citenamefont {Segovia}, \citenamefont {Garnier},\ and\
  \citenamefont {Baer}}]{Hengsberger.1999b}%
  \BibitemOpen
  \bibfield  {author} {\bibinfo {author} {\bibfnamefont {M.}~\bibnamefont
  {Hengsberger}}, \bibinfo {author} {\bibfnamefont {D.}~\bibnamefont {Purdie}},
  \bibinfo {author} {\bibfnamefont {P.}~\bibnamefont {Segovia}}, \bibinfo
  {author} {\bibfnamefont {M.}~\bibnamefont {Garnier}},\ and\ \bibinfo {author}
  {\bibfnamefont {Y.}~\bibnamefont {Baer}},\ }\bibfield  {title} {\bibinfo
  {title} {\textit{Photoemission Study of a Strongly Coupled Electron-Phonon
  System}},\ }\href {https://doi.org/10.1103/PhysRevLett.83.592} {\bibfield
  {journal} {\bibinfo  {journal} {Phys. Rev. Lett.}\ }\textbf {\bibinfo
  {volume} {83}},\ \bibinfo {pages} {592} (\bibinfo {year}
  {1999}{\natexlab{a}})}\BibitemShut {NoStop}%
\bibitem [{\citenamefont {Hengsberger}\ \emph
  {et~al.}(1999{\natexlab{b}})\citenamefont {Hengsberger}, \citenamefont
  {Fr{\'e}sard}, \citenamefont {Purdie}, \citenamefont {Segovia},\ and\
  \citenamefont {Baer}}]{Hengsberger.1999}%
  \BibitemOpen
  \bibfield  {author} {\bibinfo {author} {\bibfnamefont {M.}~\bibnamefont
  {Hengsberger}}, \bibinfo {author} {\bibfnamefont {R.}~\bibnamefont
  {Fr{\'e}sard}}, \bibinfo {author} {\bibfnamefont {D.}~\bibnamefont {Purdie}},
  \bibinfo {author} {\bibfnamefont {P.}~\bibnamefont {Segovia}},\ and\ \bibinfo
  {author} {\bibfnamefont {Y.}~\bibnamefont {Baer}},\ }\bibfield  {title}
  {\bibinfo {title} {\textit{Electron-phonon Coupling in Photoemission
  Spectra}},\ }\href {https://doi.org/10.1103/PhysRevB.60.10796} {\bibfield
  {journal} {\bibinfo  {journal} {Phys. Rev. B}\ }\textbf {\bibinfo {volume}
  {60}},\ \bibinfo {pages} {10796} (\bibinfo {year}
  {1999}{\natexlab{b}})}\BibitemShut {NoStop}%
\bibitem [{\citenamefont {Rotenberg}\ \emph {et~al.}(2000)\citenamefont
  {Rotenberg}, \citenamefont {Schaefer},\ and\ \citenamefont
  {Kevan}}]{Rotenberg.2000}%
  \BibitemOpen
  \bibfield  {author} {\bibinfo {author} {\bibnamefont {Rotenberg}}, \bibinfo
  {author} {\bibnamefont {Schaefer}},\ and\ \bibinfo {author} {\bibnamefont
  {Kevan}},\ }\bibfield  {title} {\bibinfo {title} {\textit{Coupling Between
  Adsorbate Vibrations and an Electronic Surface State}},\ }\href
  {https://doi.org/10.1103/PhysRevLett.84.2925} {\bibfield  {journal} {\bibinfo
   {journal} {Phys. Rev. Lett.}\ }\textbf {\bibinfo {volume} {84}},\ \bibinfo
  {pages} {2925} (\bibinfo {year} {2000})}\BibitemShut {NoStop}%
\bibitem [{\citenamefont {Kondo}\ \emph {et~al.}(2013)\citenamefont {Kondo},
  \citenamefont {Nakashima}, \citenamefont {Ota}, \citenamefont {Ishida},
  \citenamefont {Malaeb}, \citenamefont {Okazaki}, \citenamefont {Shin},
  \citenamefont {Kriener}, \citenamefont {Sasaki}, \citenamefont {Segawa},\
  and\ \citenamefont {Ando}}]{Kondo.2013}%
  \BibitemOpen
  \bibfield  {author} {\bibinfo {author} {\bibfnamefont {T.}~\bibnamefont
  {Kondo}}, \bibinfo {author} {\bibfnamefont {Y.}~\bibnamefont {Nakashima}},
  \bibinfo {author} {\bibfnamefont {Y.}~\bibnamefont {Ota}}, \bibinfo {author}
  {\bibfnamefont {Y.}~\bibnamefont {Ishida}}, \bibinfo {author} {\bibfnamefont
  {W.}~\bibnamefont {Malaeb}}, \bibinfo {author} {\bibfnamefont
  {K.}~\bibnamefont {Okazaki}}, \bibinfo {author} {\bibfnamefont
  {S.}~\bibnamefont {Shin}}, \bibinfo {author} {\bibfnamefont {M.}~\bibnamefont
  {Kriener}}, \bibinfo {author} {\bibfnamefont {S.}~\bibnamefont {Sasaki}},
  \bibinfo {author} {\bibfnamefont {K.}~\bibnamefont {Segawa}},\ and\ \bibinfo
  {author} {\bibfnamefont {Y.}~\bibnamefont {Ando}},\ }\bibfield  {title}
  {\bibinfo {title} {\textit{Anomalous Dressing of Dirac Fermions in the
  Topological Surface State of ${ Bi_2Se_3}$, ${ Bi_2Te_3}$, and Cu-doped ${
  Bi_2Se_3}$}},\ }\href {https://doi.org/10.1103/PhysRevLett.110.217601}
  {\bibfield  {journal} {\bibinfo  {journal} {Phys. Rev. Lett.}\ }\textbf
  {\bibinfo {volume} {110}},\ \bibinfo {pages} {217601} (\bibinfo {year}
  {2013})}\BibitemShut {NoStop}%
\bibitem [{\citenamefont {Yang}\ \emph {et~al.}(2019)\citenamefont {Yang},
  \citenamefont {Sobota}, \citenamefont {He}, \citenamefont {Leuenberger},
  \citenamefont {Soifer}, \citenamefont {Eisaki}, \citenamefont {Kirchmann},\
  and\ \citenamefont {Shen}}]{Yang.2019}%
  \BibitemOpen
  \bibfield  {author} {\bibinfo {author} {\bibfnamefont {S.-L.}\ \bibnamefont
  {Yang}}, \bibinfo {author} {\bibfnamefont {J.~A.}\ \bibnamefont {Sobota}},
  \bibinfo {author} {\bibfnamefont {Y.}~\bibnamefont {He}}, \bibinfo {author}
  {\bibfnamefont {D.}~\bibnamefont {Leuenberger}}, \bibinfo {author}
  {\bibfnamefont {H.}~\bibnamefont {Soifer}}, \bibinfo {author} {\bibfnamefont
  {H.}~\bibnamefont {Eisaki}}, \bibinfo {author} {\bibfnamefont {P.~S.}\
  \bibnamefont {Kirchmann}},\ and\ \bibinfo {author} {\bibfnamefont {Z.-X.}\
  \bibnamefont {Shen}},\ }\bibfield  {title} {\bibinfo {title}
  {\textit{Mode-Selective Coupling of Coherent Phonons to the $Bi2212$
  Electronic Band Structure}},\ }\href
  {https://doi.org/10.1103/PhysRevLett.122.176403} {\bibfield  {journal}
  {\bibinfo  {journal} {Phys. Rev. Lett.}\ }\textbf {\bibinfo {volume} {122}},\
  \bibinfo {pages} {176403} (\bibinfo {year} {2019})}\BibitemShut {NoStop}%
\bibitem [{\citenamefont {Zhou}\ \emph {et~al.}(2008)\citenamefont {Zhou},
  \citenamefont {Siegel}, \citenamefont {Fedorov},\ and\ \citenamefont
  {Lanzara}}]{Zhou.2008}%
  \BibitemOpen
  \bibfield  {author} {\bibinfo {author} {\bibfnamefont {S.~Y.}\ \bibnamefont
  {Zhou}}, \bibinfo {author} {\bibfnamefont {D.~A.}\ \bibnamefont {Siegel}},
  \bibinfo {author} {\bibfnamefont {A.~V.}\ \bibnamefont {Fedorov}},\ and\
  \bibinfo {author} {\bibfnamefont {A.}~\bibnamefont {Lanzara}},\ }\bibfield
  {title} {\bibinfo {title} {\textit{Kohn Anomaly and Interplay of
  Electron-electron and Electron-phonon Interactions in Epitaxial Graphene}},\
  }\href {https://doi.org/10.1103/PhysRevB.78.193404} {\bibfield  {journal}
  {\bibinfo  {journal} {Phys. Rev. B}\ }\textbf {\bibinfo {volume} {78}},\
  \bibinfo {pages} {193404} (\bibinfo {year} {2008})}\BibitemShut {NoStop}%
\bibitem [{\citenamefont {Butler}\ \emph {et~al.}(1979)\citenamefont {Butler},
  \citenamefont {Pinski},\ and\ \citenamefont {Allen}}]{Butler.1979}%
  \BibitemOpen
  \bibfield  {author} {\bibinfo {author} {\bibfnamefont {W.~H.}\ \bibnamefont
  {Butler}}, \bibinfo {author} {\bibfnamefont {F.~J.}\ \bibnamefont {Pinski}},\
  and\ \bibinfo {author} {\bibfnamefont {P.~B.}\ \bibnamefont {Allen}},\
  }\bibfield  {title} {\bibinfo {title} {\textit{Phonon Linewidths and
  Electron-phonon Interaction in Nb}},\ }\href
  {https://doi.org/10.1103/PhysRevB.19.3708} {\bibfield  {journal} {\bibinfo
  {journal} {Phys. Rev. B}\ }\textbf {\bibinfo {volume} {19}},\ \bibinfo
  {pages} {3708} (\bibinfo {year} {1979})}\BibitemShut {NoStop}%
\bibitem [{\citenamefont {Wakabayashi}(1986)}]{Wakabayashi.1986}%
  \BibitemOpen
  \bibfield  {author} {\bibinfo {author} {\bibfnamefont {N.}~\bibnamefont
  {Wakabayashi}},\ }\bibfield  {title} {\bibinfo {title} {\textit{Phonon
  Anomalies and Linewidths in {Nb} at 10 K}},\ }\href
  {https://doi.org/10.1103/PhysRevB.33.6771} {\bibfield  {journal} {\bibinfo
  {journal} {Phys. Rev. B}\ }\textbf {\bibinfo {volume} {33}},\ \bibinfo
  {pages} {6771} (\bibinfo {year} {1986})}\BibitemShut {NoStop}%
\bibitem [{\citenamefont {Kohn}(1959)}]{Kohn.1959}%
  \BibitemOpen
  \bibfield  {author} {\bibinfo {author} {\bibfnamefont {W.}~\bibnamefont
  {Kohn}},\ }\bibfield  {title} {\bibinfo {title} {\textit{Image of the Fermi
  Surface in the Vibration Spectrum of a Metal}},\ }\href
  {https://doi.org/10.1103/PhysRevLett.2.393} {\bibfield  {journal} {\bibinfo
  {journal} {Phys. Rev. Lett.}\ }\textbf {\bibinfo {volume} {2}},\ \bibinfo
  {pages} {393} (\bibinfo {year} {1959})}\BibitemShut {NoStop}%
\bibitem [{\citenamefont {Woll}\ and\ \citenamefont {Kohn}(1962)}]{Kohn.1962}%
  \BibitemOpen
  \bibfield  {author} {\bibinfo {author} {\bibfnamefont {E.~J.}\ \bibnamefont
  {Woll}}\ and\ \bibinfo {author} {\bibfnamefont {W.}~\bibnamefont {Kohn}},\
  }\bibfield  {title} {\bibinfo {title} {\textit{Images of the Fermi Surface in
  Phonon Spectra of Metals}},\ }\href
  {https://doi.org/10.1103/PhysRev.126.1693} {\bibfield  {journal} {\bibinfo
  {journal} {Phys. Rev.}\ }\textbf {\bibinfo {volume} {126}},\ \bibinfo {pages}
  {1693} (\bibinfo {year} {1962})}\BibitemShut {NoStop}%
\bibitem [{\citenamefont {Renker}\ \emph {et~al.}(1973)\citenamefont {Renker},
  \citenamefont {Rietschel}, \citenamefont {Pintschovius}, \citenamefont
  {Gl{\"a}ser}, \citenamefont {Br{\"u}esch}, \citenamefont {Kuse},\ and\
  \citenamefont {Rice}}]{Renker.1973}%
  \BibitemOpen
  \bibfield  {author} {\bibinfo {author} {\bibfnamefont {B.}~\bibnamefont
  {Renker}}, \bibinfo {author} {\bibfnamefont {H.}~\bibnamefont {Rietschel}},
  \bibinfo {author} {\bibfnamefont {L.}~\bibnamefont {Pintschovius}}, \bibinfo
  {author} {\bibfnamefont {W.}~\bibnamefont {Gl{\"a}ser}}, \bibinfo {author}
  {\bibfnamefont {P.}~\bibnamefont {Br{\"u}esch}}, \bibinfo {author}
  {\bibfnamefont {D.}~\bibnamefont {Kuse}},\ and\ \bibinfo {author}
  {\bibfnamefont {M.~J.}\ \bibnamefont {Rice}},\ }\bibfield  {title} {\bibinfo
  {title} {\textit{Observation of Giant Kohn Anomaly in the One-Dimensional
  Conductor ${ K_2Pt(CN)_4Br_{0.3}\cdot 3H_2O}$}},\ }\href
  {https://doi.org/10.1103/PhysRevLett.30.1144} {\bibfield  {journal} {\bibinfo
   {journal} {Phys. Rev. Lett.}\ }\textbf {\bibinfo {volume} {30}},\ \bibinfo
  {pages} {1144} (\bibinfo {year} {1973})}\BibitemShut {NoStop}%
\bibitem [{\citenamefont {Shirane}\ \emph {et~al.}(1976)\citenamefont
  {Shirane}, \citenamefont {Shapiro}, \citenamefont {Com{\`e}s}, \citenamefont
  {Garito},\ and\ \citenamefont {Heeger}}]{Shirane.1976}%
  \BibitemOpen
  \bibfield  {author} {\bibinfo {author} {\bibfnamefont {G.}~\bibnamefont
  {Shirane}}, \bibinfo {author} {\bibfnamefont {S.~M.}\ \bibnamefont
  {Shapiro}}, \bibinfo {author} {\bibfnamefont {R.}~\bibnamefont {Com{\`e}s}},
  \bibinfo {author} {\bibfnamefont {A.~F.}\ \bibnamefont {Garito}},\ and\
  \bibinfo {author} {\bibfnamefont {A.~J.}\ \bibnamefont {Heeger}},\ }\bibfield
   {title} {\bibinfo {title} {\textit{Phonon Dispersion and Kohn Anomaly in
  Tetrathiafulvalene-tetracyanoquinodimethane (TTF-TCNQ)}},\ }\href
  {https://doi.org/10.1103/PhysRevB.14.2325} {\bibfield  {journal} {\bibinfo
  {journal} {Phys. Rev. B}\ }\textbf {\bibinfo {volume} {14}},\ \bibinfo
  {pages} {2325} (\bibinfo {year} {1976})}\BibitemShut {NoStop}%
\bibitem [{\citenamefont {Pouget}\ \emph {et~al.}(1991)\citenamefont {Pouget},
  \citenamefont {Hennion}, \citenamefont {Escribe-Filippini},\ and\
  \citenamefont {Sato}}]{Pouget.1991}%
  \BibitemOpen
  \bibfield  {author} {\bibinfo {author} {\bibfnamefont {J.~P.}\ \bibnamefont
  {Pouget}}, \bibinfo {author} {\bibfnamefont {B.}~\bibnamefont {Hennion}},
  \bibinfo {author} {\bibfnamefont {C.}~\bibnamefont {Escribe-Filippini}},\
  and\ \bibinfo {author} {\bibfnamefont {M.}~\bibnamefont {Sato}},\ }\bibfield
  {title} {\bibinfo {title} {\textit{Neutron-scattering Investigations of the
  Kohn Anomaly and of the Phase and Amplitude Charge-density-wave Excitations
  of the Blue Bronze ${ K_{0.3}MoO_3}$}},\ }\href
  {https://doi.org/10.1103/PhysRevB.43.8421} {\bibfield  {journal} {\bibinfo
  {journal} {Phys. Rev. B}\ }\textbf {\bibinfo {volume} {43}},\ \bibinfo
  {pages} {8421} (\bibinfo {year} {1991})}\BibitemShut {NoStop}%
\bibitem [{\citenamefont {Guster}\ \emph {et~al.}(2019)\citenamefont {Guster},
  \citenamefont {Pruneda}, \citenamefont {Ordej{\'o}n}, \citenamefont
  {Canadell},\ and\ \citenamefont {Pouget}}]{Guster.2019}%
  \BibitemOpen
  \bibfield  {author} {\bibinfo {author} {\bibfnamefont {B.}~\bibnamefont
  {Guster}}, \bibinfo {author} {\bibfnamefont {M.}~\bibnamefont {Pruneda}},
  \bibinfo {author} {\bibfnamefont {P.}~\bibnamefont {Ordej{\'o}n}}, \bibinfo
  {author} {\bibfnamefont {E.}~\bibnamefont {Canadell}},\ and\ \bibinfo
  {author} {\bibfnamefont {J.-P.}\ \bibnamefont {Pouget}},\ }\bibfield  {title}
  {\bibinfo {title} {\textit{Evidence for the Weak Coupling Scenario of the
  Peierls Transition in the Blue Bronze}},\ }\href
  {https://doi.org/10.1103/PhysRevMaterials.3.055001} {\bibfield  {journal}
  {\bibinfo  {journal} {Phys. Rev. Mater.}\ }\textbf {\bibinfo {volume} {3}},\
  \bibinfo {pages} {055001} (\bibinfo {year} {2019})}\BibitemShut {NoStop}%
\bibitem [{\citenamefont {Politano}\ \emph {et~al.}(2015)\citenamefont
  {Politano}, \citenamefont {de~Juan}, \citenamefont {Chiarello},\ and\
  \citenamefont {Fertig}}]{Politano.2015}%
  \BibitemOpen
  \bibfield  {author} {\bibinfo {author} {\bibfnamefont {A.}~\bibnamefont
  {Politano}}, \bibinfo {author} {\bibfnamefont {F.}~\bibnamefont {de~Juan}},
  \bibinfo {author} {\bibfnamefont {G.}~\bibnamefont {Chiarello}},\ and\
  \bibinfo {author} {\bibfnamefont {H.~A.}\ \bibnamefont {Fertig}},\ }\bibfield
   {title} {\bibinfo {title} {\textit{Emergence of an Out-of-Plane Optical
  Phonon {(ZO)} Kohn Anomaly in Quasifreestanding Epitaxial Graphene}},\ }\href
  {https://doi.org/10.1103/PhysRevLett.115.075504} {\bibfield  {journal}
  {\bibinfo  {journal} {Phys. Rev. Lett.}\ }\textbf {\bibinfo {volume} {115}},\
  \bibinfo {pages} {075504} (\bibinfo {year} {2015})}\BibitemShut {NoStop}%
\bibitem [{\citenamefont {Haddad}\ and\ \citenamefont
  {Mandhour}(2018)}]{Haddad.2018}%
  \BibitemOpen
  \bibfield  {author} {\bibinfo {author} {\bibfnamefont {S.}~\bibnamefont
  {Haddad}}\ and\ \bibinfo {author} {\bibfnamefont {L.}~\bibnamefont
  {Mandhour}},\ }\bibfield  {title} {\bibinfo {title} {\textit{Kohn Anomaly of
  Optical Zone Boundary Phonons in Uniaxial Strained Graphene: {Role} of the
  Dirac Cone Electronic Dispersion}},\ }\href
  {https://doi.org/10.1103/PhysRevB.98.115420} {\bibfield  {journal} {\bibinfo
  {journal} {Phys. Rev. B}\ }\textbf {\bibinfo {volume} {98}},\ \bibinfo
  {pages} {115420} (\bibinfo {year} {2018})}\BibitemShut {NoStop}%
\bibitem [{\citenamefont {Weber}\ \emph {et~al.}(2011)\citenamefont {Weber},
  \citenamefont {Rosenkranz}, \citenamefont {Castellan}, \citenamefont
  {Osborn}, \citenamefont {Hott}, \citenamefont {Heid}, \citenamefont {Bohnen},
  \citenamefont {Egami}, \citenamefont {Said},\ and\ \citenamefont
  {Reznik}}]{Weber.2011}%
  \BibitemOpen
  \bibfield  {author} {\bibinfo {author} {\bibfnamefont {F.}~\bibnamefont
  {Weber}}, \bibinfo {author} {\bibfnamefont {S.}~\bibnamefont {Rosenkranz}},
  \bibinfo {author} {\bibfnamefont {J.-P.}\ \bibnamefont {Castellan}}, \bibinfo
  {author} {\bibfnamefont {R.}~\bibnamefont {Osborn}}, \bibinfo {author}
  {\bibfnamefont {R.}~\bibnamefont {Hott}}, \bibinfo {author} {\bibfnamefont
  {R.}~\bibnamefont {Heid}}, \bibinfo {author} {\bibfnamefont {K.-P.}\
  \bibnamefont {Bohnen}}, \bibinfo {author} {\bibfnamefont {T.}~\bibnamefont
  {Egami}}, \bibinfo {author} {\bibfnamefont {A.~H.}\ \bibnamefont {Said}},\
  and\ \bibinfo {author} {\bibfnamefont {D.}~\bibnamefont {Reznik}},\
  }\bibfield  {title} {\bibinfo {title} {\textit{Extended Phonon Collapse and
  the Origin of the Charge-Density Wave in
  $2H\mathrm{\text{\ensuremath{-}}}{\mathrm{NbSe}}_{2}$}},\ }\href
  {https://doi.org/10.1103/PhysRevLett.107.107403} {\bibfield  {journal}
  {\bibinfo  {journal} {Phys. Rev. Lett.}\ }\textbf {\bibinfo {volume} {107}},\
  \bibinfo {pages} {107403} (\bibinfo {year} {2011})}\BibitemShut {NoStop}%
\bibitem [{\citenamefont {Brockhouse}\ \emph {et~al.}(1962)\citenamefont
  {Brockhouse}, \citenamefont {Arase}, \citenamefont {Caglioti}, \citenamefont
  {Rao},\ and\ \citenamefont {Woods}}]{Brockhouse.1962}%
  \BibitemOpen
  \bibfield  {author} {\bibinfo {author} {\bibfnamefont {B.~N.}\ \bibnamefont
  {Brockhouse}}, \bibinfo {author} {\bibfnamefont {T.}~\bibnamefont {Arase}},
  \bibinfo {author} {\bibfnamefont {G.}~\bibnamefont {Caglioti}}, \bibinfo
  {author} {\bibfnamefont {K.~R.}\ \bibnamefont {Rao}},\ and\ \bibinfo {author}
  {\bibfnamefont {A.~D.~B.}\ \bibnamefont {Woods}},\ }\bibfield  {title}
  {\bibinfo {title} {\textit{Crystal Dynamics of Lead. I. Dispersion Curves at
  $100^\circ$K}},\ }\href {https://doi.org/10.1103/PhysRev.128.1099} {\bibfield
   {journal} {\bibinfo  {journal} {Phys. Rev.}\ }\textbf {\bibinfo {volume}
  {128}},\ \bibinfo {pages} {1099} (\bibinfo {year} {1962})}\BibitemShut
  {NoStop}%
\bibitem [{\citenamefont {Wong}\ \emph {et~al.}(2003)\citenamefont {Wong},
  \citenamefont {Krisch}, \citenamefont {Farber}, \citenamefont {Occelli},
  \citenamefont {Schwartz}, \citenamefont {Chiang}, \citenamefont {Wall},
  \citenamefont {Boro},\ and\ \citenamefont {Xu}}]{Wong.2003}%
  \BibitemOpen
  \bibfield  {author} {\bibinfo {author} {\bibfnamefont {J.}~\bibnamefont
  {Wong}}, \bibinfo {author} {\bibfnamefont {M.}~\bibnamefont {Krisch}},
  \bibinfo {author} {\bibfnamefont {D.~L.}\ \bibnamefont {Farber}}, \bibinfo
  {author} {\bibfnamefont {F.}~\bibnamefont {Occelli}}, \bibinfo {author}
  {\bibfnamefont {A.~J.}\ \bibnamefont {Schwartz}}, \bibinfo {author}
  {\bibfnamefont {T.-C.}\ \bibnamefont {Chiang}}, \bibinfo {author}
  {\bibfnamefont {M.}~\bibnamefont {Wall}}, \bibinfo {author} {\bibfnamefont
  {C.}~\bibnamefont {Boro}},\ and\ \bibinfo {author} {\bibfnamefont
  {R.}~\bibnamefont {Xu}},\ }\bibfield  {title} {\bibinfo {title}
  {\textit{Phonon Dispersions of FCC Delta-plutonium-gallium by Inelastic x-ray
  Scattering}},\ }\href {https://doi.org/10.1126/science.1087179} {\bibfield
  {journal} {\bibinfo  {journal} {Science}\ }\textbf {\bibinfo {volume}
  {301}},\ \bibinfo {pages} {1078} (\bibinfo {year} {2003})}\BibitemShut
  {NoStop}%
\bibitem [{\citenamefont {Agrestini}\ \emph {et~al.}(2004)\citenamefont
  {Agrestini}, \citenamefont {Metallo}, \citenamefont {Filippi}, \citenamefont
  {Simonelli}, \citenamefont {Campi}, \citenamefont {Sanipoli}, \citenamefont
  {Liarokapis}, \citenamefont {De~Negri}, \citenamefont {Giovannini},
  \citenamefont {Saccone}, \citenamefont {Latini},\ and\ \citenamefont
  {Bianconi}}]{Agrestini.2004}%
  \BibitemOpen
  \bibfield  {author} {\bibinfo {author} {\bibfnamefont {S.}~\bibnamefont
  {Agrestini}}, \bibinfo {author} {\bibfnamefont {C.}~\bibnamefont {Metallo}},
  \bibinfo {author} {\bibfnamefont {M.}~\bibnamefont {Filippi}}, \bibinfo
  {author} {\bibfnamefont {L.}~\bibnamefont {Simonelli}}, \bibinfo {author}
  {\bibfnamefont {G.}~\bibnamefont {Campi}}, \bibinfo {author} {\bibfnamefont
  {C.}~\bibnamefont {Sanipoli}}, \bibinfo {author} {\bibfnamefont
  {E.}~\bibnamefont {Liarokapis}}, \bibinfo {author} {\bibfnamefont
  {S.}~\bibnamefont {De~Negri}}, \bibinfo {author} {\bibfnamefont
  {M.}~\bibnamefont {Giovannini}}, \bibinfo {author} {\bibfnamefont
  {A.}~\bibnamefont {Saccone}}, \bibinfo {author} {\bibfnamefont
  {A.}~\bibnamefont {Latini}},\ and\ \bibinfo {author} {\bibfnamefont
  {A.}~\bibnamefont {Bianconi}},\ }\bibfield  {title} {\bibinfo {title}
  {\textit{Substitution of {Sc} for {Mg} in {${ MgB_2}$}: Effects on Transition
  Temperature and Kohn Anomaly}},\ }\href
  {https://doi.org/10.1103/PhysRevB.70.134514} {\bibfield  {journal} {\bibinfo
  {journal} {Phys. Rev. B}\ }\textbf {\bibinfo {volume} {70}},\ \bibinfo
  {pages} {134514} (\bibinfo {year} {2004})}\BibitemShut {NoStop}%
\bibitem [{\citenamefont {Baron}\ \emph {et~al.}(2004)\citenamefont {Baron},
  \citenamefont {Uchiyama}, \citenamefont {Tanaka}, \citenamefont {Tsutsui},
  \citenamefont {Ishikawa}, \citenamefont {Lee}, \citenamefont {Heid},
  \citenamefont {Bohnen}, \citenamefont {Tajima},\ and\ \citenamefont
  {Ishikawa}}]{Baron.2004}%
  \BibitemOpen
  \bibfield  {author} {\bibinfo {author} {\bibfnamefont {A.~Q.~R.}\
  \bibnamefont {Baron}}, \bibinfo {author} {\bibfnamefont {H.}~\bibnamefont
  {Uchiyama}}, \bibinfo {author} {\bibfnamefont {Y.}~\bibnamefont {Tanaka}},
  \bibinfo {author} {\bibfnamefont {S.}~\bibnamefont {Tsutsui}}, \bibinfo
  {author} {\bibfnamefont {D.}~\bibnamefont {Ishikawa}}, \bibinfo {author}
  {\bibfnamefont {S.}~\bibnamefont {Lee}}, \bibinfo {author} {\bibfnamefont
  {R.}~\bibnamefont {Heid}}, \bibinfo {author} {\bibfnamefont {K.-P.}\
  \bibnamefont {Bohnen}}, \bibinfo {author} {\bibfnamefont {S.}~\bibnamefont
  {Tajima}},\ and\ \bibinfo {author} {\bibfnamefont {T.}~\bibnamefont
  {Ishikawa}},\ }\bibfield  {title} {\bibinfo {title} {\textit{Kohn Anomaly in
  ${ MgB_2}$ by Inelastic X-ray Scattering}},\ }\href
  {https://doi.org/10.1103/PhysRevLett.92.197004} {\bibfield  {journal}
  {\bibinfo  {journal} {Phys. Rev. Lett.}\ }\textbf {\bibinfo {volume} {92}},\
  \bibinfo {pages} {197004} (\bibinfo {year} {2004})}\BibitemShut {NoStop}%
\bibitem [{\citenamefont {Kaden}\ \emph {et~al.}(1992)\citenamefont {Kaden},
  \citenamefont {Ruggerone}, \citenamefont {Toennies}, \citenamefont {Zhang},\
  and\ \citenamefont {Benedek}}]{Kaden.1992}%
  \BibitemOpen
  \bibfield  {author} {\bibinfo {author} {\bibfnamefont {C.}~\bibnamefont
  {Kaden}}, \bibinfo {author} {\bibfnamefont {P.}~\bibnamefont {Ruggerone}},
  \bibinfo {author} {\bibfnamefont {J.~P.}\ \bibnamefont {Toennies}}, \bibinfo
  {author} {\bibfnamefont {G.}~\bibnamefont {Zhang}},\ and\ \bibinfo {author}
  {\bibfnamefont {G.}~\bibnamefont {Benedek}},\ }\bibfield  {title} {\bibinfo
  {title} {\textit{Electronic Pseudocharge Model for the Cu(111)
  Longitudinal-surface-phonon Anomaly Observed by Helium-atom Scattering}},\
  }\href {https://doi.org/10.1103/PhysRevB.46.13509} {\bibfield  {journal}
  {\bibinfo  {journal} {Phys. Rev. B}\ }\textbf {\bibinfo {volume} {46}},\
  \bibinfo {pages} {13509} (\bibinfo {year} {1992})}\BibitemShut {NoStop}%
\bibitem [{\citenamefont {Doak}\ \emph {et~al.}(1983)\citenamefont {Doak},
  \citenamefont {Harten},\ and\ \citenamefont {Toennies}}]{Doak.1983}%
  \BibitemOpen
  \bibfield  {author} {\bibinfo {author} {\bibfnamefont {R.~B.}\ \bibnamefont
  {Doak}}, \bibinfo {author} {\bibfnamefont {U.}~\bibnamefont {Harten}},\ and\
  \bibinfo {author} {\bibfnamefont {J.~P.}\ \bibnamefont {Toennies}},\
  }\bibfield  {title} {\bibinfo {title} {\textit{Anomalous Surface Phonon
  Dispersion Relations for Ag(111) Measured by Inelastic Scattering of He
  Atoms}},\ }\href {https://doi.org/10.1103/PhysRevLett.51.578} {\bibfield
  {journal} {\bibinfo  {journal} {Phys. Rev. Lett.}\ }\textbf {\bibinfo
  {volume} {51}},\ \bibinfo {pages} {578} (\bibinfo {year} {1983})}\BibitemShut
  {NoStop}%
\bibitem [{\citenamefont {Harten}\ \emph
  {et~al.}(1985{\natexlab{a}})\citenamefont {Harten}, \citenamefont {Toennies},
  \citenamefont {W{\"o}ll},\ and\ \citenamefont {Zhang}}]{Harten.1985}%
  \BibitemOpen
  \bibfield  {author} {\bibinfo {author} {\bibnamefont {Harten}}, \bibinfo
  {author} {\bibnamefont {Toennies}}, \bibinfo {author} {\bibnamefont
  {W{\"o}ll}},\ and\ \bibinfo {author} {\bibnamefont {Zhang}},\ }\bibfield
  {title} {\bibinfo {title} {\textit{Observation of a Kohn Anomaly in the
  Surface-phonon Dispersion Curves of Pt(111)}},\ }\href
  {https://doi.org/10.1103/PhysRevLett.55.2308} {\bibfield  {journal} {\bibinfo
   {journal} {Phys. Rev. Lett.}\ }\textbf {\bibinfo {volume} {55}},\ \bibinfo
  {pages} {2308} (\bibinfo {year} {1985}{\natexlab{a}})}\BibitemShut {NoStop}%
\bibitem [{\citenamefont {Harten}\ \emph
  {et~al.}(1985{\natexlab{b}})\citenamefont {Harten}, \citenamefont
  {Toennies},\ and\ \citenamefont {W{\"o}ll}}]{Harten.1985b}%
  \BibitemOpen
  \bibfield  {author} {\bibinfo {author} {\bibfnamefont {U.}~\bibnamefont
  {Harten}}, \bibinfo {author} {\bibfnamefont {J.~P.}\ \bibnamefont
  {Toennies}},\ and\ \bibinfo {author} {\bibfnamefont {C.}~\bibnamefont
  {W{\"o}ll}},\ }\bibfield  {title} {\bibinfo {title} {\textit{Helium
  Time-of-flight Spectroscopy of Surface-phonon Dispersion Curves of the Noble
  Metals}},\ }\href {https://doi.org/10.1039/DC9858000137} {\bibfield
  {journal} {\bibinfo  {journal} {Faraday Discuss. Chem. Soc.}\ }\textbf
  {\bibinfo {volume} {80}},\ \bibinfo {pages} {137} (\bibinfo {year}
  {1985}{\natexlab{b}})}\BibitemShut {NoStop}%
\bibitem [{\citenamefont {Jayanthi}\ \emph {et~al.}(1987)\citenamefont
  {Jayanthi}, \citenamefont {Bilz}, \citenamefont {Kress},\ and\ \citenamefont
  {Benedek}}]{Jayanthi.1987}%
  \BibitemOpen
  \bibfield  {author} {\bibinfo {author} {\bibnamefont {Jayanthi}}, \bibinfo
  {author} {\bibnamefont {Bilz}}, \bibinfo {author} {\bibnamefont {Kress}},\
  and\ \bibinfo {author} {\bibnamefont {Benedek}},\ }\bibfield  {title}
  {\bibinfo {title} {\textit{Nature of Surface-phonon Anomalies in Noble
  Metals}},\ }\href {https://doi.org/10.1103/PhysRevLett.59.795} {\bibfield
  {journal} {\bibinfo  {journal} {Phys. Rev. Lett.}\ }\textbf {\bibinfo
  {volume} {59}},\ \bibinfo {pages} {795} (\bibinfo {year} {1987})}\BibitemShut
  {NoStop}%
\bibitem [{\citenamefont {Kr{\"o}ger}\ \emph {et~al.}(1997)\citenamefont
  {Kr{\"o}ger}, \citenamefont {Lehwald},\ and\ \citenamefont
  {Ibach}}]{Kroger.1997}%
  \BibitemOpen
  \bibfield  {author} {\bibinfo {author} {\bibfnamefont {J.}~\bibnamefont
  {Kr{\"o}ger}}, \bibinfo {author} {\bibfnamefont {S.}~\bibnamefont
  {Lehwald}},\ and\ \bibinfo {author} {\bibfnamefont {H.}~\bibnamefont
  {Ibach}},\ }\bibfield  {title} {\bibinfo {title} {\textit{EELS study of the
  Clean and Hydrogen-covered Mo(110) Surface}},\ }\href
  {https://doi.org/10.1103/PhysRevB.55.10895} {\bibfield  {journal} {\bibinfo
  {journal} {Phys. Rev. B}\ }\textbf {\bibinfo {volume} {55}},\ \bibinfo
  {pages} {10895} (\bibinfo {year} {1997})}\BibitemShut {NoStop}%
\bibitem [{\citenamefont {Hulpke}\ and\ \citenamefont
  {L{\"u}decke}(1993)}]{Hulpke.1993}%
  \BibitemOpen
  \bibfield  {author} {\bibinfo {author} {\bibfnamefont {E.}~\bibnamefont
  {Hulpke}}\ and\ \bibinfo {author} {\bibfnamefont {J.}~\bibnamefont
  {L{\"u}decke}},\ }\bibfield  {title} {\bibinfo {title} {\textit{The Giant
  Surface Phonon Anomaly on Hydrogen Saturated W(110) and Mo(110)}},\ }\href
  {https://doi.org/10.1016/0039-6028(93)91083-2} {\bibfield  {journal}
  {\bibinfo  {journal} {Surf. Sci.}\ }\textbf {\bibinfo {volume} {287-288}},\
  \bibinfo {pages} {837} (\bibinfo {year} {1993})}\BibitemShut {NoStop}%
\bibitem [{\citenamefont {Kohler}\ \emph {et~al.}(1995)\citenamefont {Kohler},
  \citenamefont {Ruggerone}, \citenamefont {Wilke},\ and\ \citenamefont
  {Scheffler}}]{Kohler.1995}%
  \BibitemOpen
  \bibfield  {author} {\bibinfo {author} {\bibnamefont {Kohler}}, \bibinfo
  {author} {\bibnamefont {Ruggerone}}, \bibinfo {author} {\bibnamefont
  {Wilke}},\ and\ \bibinfo {author} {\bibnamefont {Scheffler}},\ }\bibfield
  {title} {\bibinfo {title} {\textit{Frustrated H-induced Instability of
  Mo(110)}},\ }\href {https://doi.org/10.1103/PhysRevLett.74.1387} {\bibfield
  {journal} {\bibinfo  {journal} {Phys. Rev. Lett.}\ }\textbf {\bibinfo
  {volume} {74}},\ \bibinfo {pages} {1387} (\bibinfo {year}
  {1995})}\BibitemShut {NoStop}%
\bibitem [{\citenamefont {Hulpke}\ and\ \citenamefont
  {L{\"u}decke}(1992)}]{Hulpke.1992}%
  \BibitemOpen
  \bibfield  {author} {\bibinfo {author} {\bibnamefont {Hulpke}}\ and\ \bibinfo
  {author} {\bibnamefont {L{\"u}decke}},\ }\bibfield  {title} {\bibinfo {title}
  {\textit{Hydrogen-induced Phonon Anomaly on the W(110) Surface}},\ }\href
  {https://doi.org/10.1103/PhysRevLett.68.2846} {\bibfield  {journal} {\bibinfo
   {journal} {Phys. Rev. Lett.}\ }\textbf {\bibinfo {volume} {68}},\ \bibinfo
  {pages} {2846} (\bibinfo {year} {1992})}\BibitemShut {NoStop}%
\bibitem [{\citenamefont {Rotenberg}\ and\ \citenamefont
  {Kevan}(1998)}]{Rotenberg.1998}%
  \BibitemOpen
  \bibfield  {author} {\bibinfo {author} {\bibfnamefont {E.}~\bibnamefont
  {Rotenberg}}\ and\ \bibinfo {author} {\bibfnamefont {S.~D.}\ \bibnamefont
  {Kevan}},\ }\bibfield  {title} {\bibinfo {title} {\textit{Evolution of Fermi
  Level Crossings versus H Coverage on W(110)}},\ }\href
  {https://doi.org/10.1103/PhysRevLett.80.2905} {\bibfield  {journal} {\bibinfo
   {journal} {Phys. Rev. Lett.}\ }\textbf {\bibinfo {volume} {80}},\ \bibinfo
  {pages} {2905} (\bibinfo {year} {1998})}\BibitemShut {NoStop}%
\bibitem [{\citenamefont {Benedek}\ and\ \citenamefont
  {Toennies}(2018)}]{Benedek.2018b}%
  \BibitemOpen
  \bibfield  {author} {\bibinfo {author} {\bibfnamefont {G.}~\bibnamefont
  {Benedek}}\ and\ \bibinfo {author} {\bibfnamefont {J.~P.}\ \bibnamefont
  {Toennies}},\ }\href {https://doi.org/10.1007/978-3-662-56443-1} {\emph
  {\bibinfo {title} {\textit{Atomic Scale Dynamics at Surfaces}}}},\
  Vol.~\bibinfo {volume} {63}\ (\bibinfo  {publisher} {{Springer Berlin
  Heidelberg}},\ \bibinfo {address} {Berlin, Heidelberg},\ \bibinfo {year}
  {2018})\BibitemShut {NoStop}%
\bibitem [{\citenamefont {Zhu}\ \emph {et~al.}(2012)\citenamefont {Zhu},
  \citenamefont {Santos}, \citenamefont {Howard}, \citenamefont {Sankar},
  \citenamefont {Chou}, \citenamefont {Chamon},\ and\ \citenamefont
  {El-Batanouny}}]{Zhu.2012}%
  \BibitemOpen
  \bibfield  {author} {\bibinfo {author} {\bibfnamefont {X.}~\bibnamefont
  {Zhu}}, \bibinfo {author} {\bibfnamefont {L.}~\bibnamefont {Santos}},
  \bibinfo {author} {\bibfnamefont {C.}~\bibnamefont {Howard}}, \bibinfo
  {author} {\bibfnamefont {R.}~\bibnamefont {Sankar}}, \bibinfo {author}
  {\bibfnamefont {F.~C.}\ \bibnamefont {Chou}}, \bibinfo {author}
  {\bibfnamefont {C.}~\bibnamefont {Chamon}},\ and\ \bibinfo {author}
  {\bibfnamefont {M.}~\bibnamefont {El-Batanouny}},\ }\bibfield  {title}
  {\bibinfo {title} {\textit{Electron-phonon Coupling on the Surface of the
  Topological Insulator ${ Bi_2Se_3}$ Determined from Surface-phonon Dispersion
  Measurements}},\ }\href {https://doi.org/10.1103/PhysRevLett.108.185501}
  {\bibfield  {journal} {\bibinfo  {journal} {Phys. Rev. Lett.}\ }\textbf
  {\bibinfo {volume} {108}},\ \bibinfo {pages} {185501} (\bibinfo {year}
  {2012})}\BibitemShut {NoStop}%
\bibitem [{\citenamefont {Zhu}\ \emph {et~al.}(2011)\citenamefont {Zhu},
  \citenamefont {Santos}, \citenamefont {Sankar}, \citenamefont {Chikara},
  \citenamefont {Howard}, \citenamefont {Chou}, \citenamefont {Chamon},\ and\
  \citenamefont {El-Batanouny}}]{Zhu.2011}%
  \BibitemOpen
  \bibfield  {author} {\bibinfo {author} {\bibfnamefont {X.}~\bibnamefont
  {Zhu}}, \bibinfo {author} {\bibfnamefont {L.}~\bibnamefont {Santos}},
  \bibinfo {author} {\bibfnamefont {R.}~\bibnamefont {Sankar}}, \bibinfo
  {author} {\bibfnamefont {S.}~\bibnamefont {Chikara}}, \bibinfo {author}
  {\bibfnamefont {C.}~\bibnamefont {Howard}}, \bibinfo {author} {\bibfnamefont
  {F.~C.}\ \bibnamefont {Chou}}, \bibinfo {author} {\bibfnamefont
  {C.}~\bibnamefont {Chamon}},\ and\ \bibinfo {author} {\bibfnamefont
  {M.}~\bibnamefont {El-Batanouny}},\ }\bibfield  {title} {\bibinfo {title}
  {\textit{Interaction of Phonons and Dirac Fermions on the Surface of ${
  Bi_2Se_3}$: A Strong Kohn Anomaly}},\ }\href
  {https://doi.org/10.1103/PhysRevLett.107.186102} {\bibfield  {journal}
  {\bibinfo  {journal} {Phys. Rev. Lett.}\ }\textbf {\bibinfo {volume} {107}},\
  \bibinfo {pages} {186102} (\bibinfo {year} {2011})}\BibitemShut {NoStop}%
\bibitem [{\citenamefont {Howard}\ \emph {et~al.}(2013)\citenamefont {Howard},
  \citenamefont {El-Batanouny}, \citenamefont {Sankar},\ and\ \citenamefont
  {Chou}}]{Howard.2013}%
  \BibitemOpen
  \bibfield  {author} {\bibinfo {author} {\bibfnamefont {C.}~\bibnamefont
  {Howard}}, \bibinfo {author} {\bibfnamefont {M.}~\bibnamefont
  {El-Batanouny}}, \bibinfo {author} {\bibfnamefont {R.}~\bibnamefont
  {Sankar}},\ and\ \bibinfo {author} {\bibfnamefont {F.~C.}\ \bibnamefont
  {Chou}},\ }\bibfield  {title} {\bibinfo {title} {\textit{Anomalous Behavior
  in the Phonon Dispersion of the (001) Surface of ${ Bi_2Te_3}$ Determined
  from Helium Atom-surface Scattering Measurements}},\ }\href
  {https://doi.org/10.1103/PhysRevB.88.035402} {\bibfield  {journal} {\bibinfo
  {journal} {Phys. Rev. B}\ }\textbf {\bibinfo {volume} {88}},\ \bibinfo
  {pages} {035402} (\bibinfo {year} {2013})}\BibitemShut {NoStop}%
\bibitem [{\citenamefont {Kalish}\ \emph {et~al.}(2019)\citenamefont {Kalish},
  \citenamefont {Chamon}, \citenamefont {El-Batanouny}, \citenamefont {Santos},
  \citenamefont {Sankar},\ and\ \citenamefont {Chou}}]{Kalish.2019}%
  \BibitemOpen
  \bibfield  {author} {\bibinfo {author} {\bibfnamefont {S.}~\bibnamefont
  {Kalish}}, \bibinfo {author} {\bibfnamefont {C.}~\bibnamefont {Chamon}},
  \bibinfo {author} {\bibfnamefont {M.}~\bibnamefont {El-Batanouny}}, \bibinfo
  {author} {\bibfnamefont {L.~H.}\ \bibnamefont {Santos}}, \bibinfo {author}
  {\bibfnamefont {R.}~\bibnamefont {Sankar}},\ and\ \bibinfo {author}
  {\bibfnamefont {F.~C.}\ \bibnamefont {Chou}},\ }\bibfield  {title} {\bibinfo
  {title} {\textit{Contrasting the Surface Phonon Dispersion of ${
  Pb_{0.7}Sn_{0.3}Se}$ in Its Topologically Trivial and Nontrivial Phases}},\
  }\href {https://doi.org/10.1103/PhysRevLett.122.116101} {\bibfield  {journal}
  {\bibinfo  {journal} {Phys. Rev. Lett.}\ }\textbf {\bibinfo {volume} {122}},\
  \bibinfo {pages} {116101} (\bibinfo {year} {2019})}\BibitemShut {NoStop}%
\bibitem [{\citenamefont {Topp}\ \emph {et~al.}(2017)\citenamefont {Topp},
  \citenamefont {Queiroz}, \citenamefont {Gr{\"u}neis}, \citenamefont
  {M{\"u}chler}, \citenamefont {Rost}, \citenamefont {Varykhalov},
  \citenamefont {Marchenko}, \citenamefont {Krivenkov}, \citenamefont
  {Rodolakis}, \citenamefont {McChesney}, \citenamefont {Lotsch}, \citenamefont
  {Schoop},\ and\ \citenamefont {Ast}}]{Topp.2017}%
  \BibitemOpen
  \bibfield  {author} {\bibinfo {author} {\bibfnamefont {A.}~\bibnamefont
  {Topp}}, \bibinfo {author} {\bibfnamefont {R.}~\bibnamefont {Queiroz}},
  \bibinfo {author} {\bibfnamefont {A.}~\bibnamefont {Gr{\"u}neis}}, \bibinfo
  {author} {\bibfnamefont {L.}~\bibnamefont {M{\"u}chler}}, \bibinfo {author}
  {\bibfnamefont {A.~W.}\ \bibnamefont {Rost}}, \bibinfo {author}
  {\bibfnamefont {A.}~\bibnamefont {Varykhalov}}, \bibinfo {author}
  {\bibfnamefont {D.}~\bibnamefont {Marchenko}}, \bibinfo {author}
  {\bibfnamefont {M.}~\bibnamefont {Krivenkov}}, \bibinfo {author}
  {\bibfnamefont {F.}~\bibnamefont {Rodolakis}}, \bibinfo {author}
  {\bibfnamefont {J.~L.}\ \bibnamefont {McChesney}}, \bibinfo {author}
  {\bibfnamefont {B.~V.}\ \bibnamefont {Lotsch}}, \bibinfo {author}
  {\bibfnamefont {L.~M.}\ \bibnamefont {Schoop}},\ and\ \bibinfo {author}
  {\bibfnamefont {C.~R.}\ \bibnamefont {Ast}},\ }\bibfield  {title} {\bibinfo
  {title} {\textit{Surface Floating 2D Bands in Layered Nonsymmorphic
  Semimetals: ZrSiS and Related Compounds}},\ }\href
  {https://doi.org/10.1103/PhysRevX.7.041073} {\bibfield  {journal} {\bibinfo
  {journal} {Phys. Rev. X}\ }\textbf {\bibinfo {volume} {7}},\ \bibinfo {pages}
  {041073} (\bibinfo {year} {2017})}\BibitemShut {NoStop}%
\bibitem [{\citenamefont {Schoop}\ \emph {et~al.}(2016)\citenamefont {Schoop},
  \citenamefont {Ali}, \citenamefont {Stra{\ss}er}, \citenamefont {Topp},
  \citenamefont {Varykhalov}, \citenamefont {Marchenko}, \citenamefont
  {Duppel}, \citenamefont {Parkin}, \citenamefont {Lotsch},\ and\ \citenamefont
  {Ast}}]{Schoop.2016}%
  \BibitemOpen
  \bibfield  {author} {\bibinfo {author} {\bibfnamefont {L.~M.}\ \bibnamefont
  {Schoop}}, \bibinfo {author} {\bibfnamefont {M.~N.}\ \bibnamefont {Ali}},
  \bibinfo {author} {\bibfnamefont {C.}~\bibnamefont {Stra{\ss}er}}, \bibinfo
  {author} {\bibfnamefont {A.}~\bibnamefont {Topp}}, \bibinfo {author}
  {\bibfnamefont {A.}~\bibnamefont {Varykhalov}}, \bibinfo {author}
  {\bibfnamefont {D.}~\bibnamefont {Marchenko}}, \bibinfo {author}
  {\bibfnamefont {V.}~\bibnamefont {Duppel}}, \bibinfo {author} {\bibfnamefont
  {S.~S.~P.}\ \bibnamefont {Parkin}}, \bibinfo {author} {\bibfnamefont {B.~V.}\
  \bibnamefont {Lotsch}},\ and\ \bibinfo {author} {\bibfnamefont {C.~R.}\
  \bibnamefont {Ast}},\ }\bibfield  {title} {\bibinfo {title} {\textit{Dirac
  Cone Protected by Non-symmorphic Symmetry and Three-dimensional Dirac Line
  Node in ZrSiS}},\ }\href {https://doi.org/10.1038/ncomms11696} {\bibfield
  {journal} {\bibinfo  {journal} {Nat. Commun.}\ }\textbf {\bibinfo {volume}
  {7}},\ \bibinfo {pages} {11696} (\bibinfo {year} {2016})}\BibitemShut
  {NoStop}%
\bibitem [{\citenamefont {Neupane}\ \emph {et~al.}(2016)\citenamefont
  {Neupane}, \citenamefont {Belopolski}, \citenamefont {Hosen}, \citenamefont
  {Sanchez}, \citenamefont {Sankar}, \citenamefont {Szlawska}, \citenamefont
  {Xu}, \citenamefont {Dimitri}, \citenamefont {Dhakal}, \citenamefont
  {Maldonado}, \citenamefont {Oppeneer}, \citenamefont {Kaczorowski},
  \citenamefont {Chou}, \citenamefont {Hasan},\ and\ \citenamefont
  {Durakiewicz}}]{Neupane.2016}%
  \BibitemOpen
  \bibfield  {author} {\bibinfo {author} {\bibfnamefont {M.}~\bibnamefont
  {Neupane}}, \bibinfo {author} {\bibfnamefont {I.}~\bibnamefont {Belopolski}},
  \bibinfo {author} {\bibfnamefont {M.~M.}\ \bibnamefont {Hosen}}, \bibinfo
  {author} {\bibfnamefont {D.~S.}\ \bibnamefont {Sanchez}}, \bibinfo {author}
  {\bibfnamefont {R.}~\bibnamefont {Sankar}}, \bibinfo {author} {\bibfnamefont
  {M.}~\bibnamefont {Szlawska}}, \bibinfo {author} {\bibfnamefont {S.-Y.}\
  \bibnamefont {Xu}}, \bibinfo {author} {\bibfnamefont {K.}~\bibnamefont
  {Dimitri}}, \bibinfo {author} {\bibfnamefont {N.}~\bibnamefont {Dhakal}},
  \bibinfo {author} {\bibfnamefont {P.}~\bibnamefont {Maldonado}}, \bibinfo
  {author} {\bibfnamefont {P.~M.}\ \bibnamefont {Oppeneer}}, \bibinfo {author}
  {\bibfnamefont {D.}~\bibnamefont {Kaczorowski}}, \bibinfo {author}
  {\bibfnamefont {F.}~\bibnamefont {Chou}}, \bibinfo {author} {\bibfnamefont
  {M.~Z.}\ \bibnamefont {Hasan}},\ and\ \bibinfo {author} {\bibfnamefont
  {T.}~\bibnamefont {Durakiewicz}},\ }\bibfield  {title} {\bibinfo {title}
  {\textit{Observation of Topological Nodal Fermion Semimetal Phase in
  Zrsis}},\ }\href {https://doi.org/10.1103/PhysRevB.93.201104} {\bibfield
  {journal} {\bibinfo  {journal} {Phys. Rev. B}\ }\textbf {\bibinfo {volume}
  {93}},\ \bibinfo {pages} {201104(R)} (\bibinfo {year} {2016})}\BibitemShut
  {NoStop}%
\bibitem [{\citenamefont {Wang}\ \emph {et~al.}(2016)\citenamefont {Wang},
  \citenamefont {Pan}, \citenamefont {Gao}, \citenamefont {Yu}, \citenamefont
  {Jiang}, \citenamefont {Zhang}, \citenamefont {Zuo}, \citenamefont {Zhang},
  \citenamefont {Wei}, \citenamefont {Niu}, \citenamefont {Xia}, \citenamefont
  {Wan}, \citenamefont {Chen}, \citenamefont {Song}, \citenamefont {Xu},
  \citenamefont {Wang}, \citenamefont {Wang},\ and\ \citenamefont
  {Zhang}}]{Wang.2016}%
  \BibitemOpen
  \bibfield  {author} {\bibinfo {author} {\bibfnamefont {X.}~\bibnamefont
  {Wang}}, \bibinfo {author} {\bibfnamefont {X.}~\bibnamefont {Pan}}, \bibinfo
  {author} {\bibfnamefont {M.}~\bibnamefont {Gao}}, \bibinfo {author}
  {\bibfnamefont {J.}~\bibnamefont {Yu}}, \bibinfo {author} {\bibfnamefont
  {J.}~\bibnamefont {Jiang}}, \bibinfo {author} {\bibfnamefont
  {J.}~\bibnamefont {Zhang}}, \bibinfo {author} {\bibfnamefont
  {H.}~\bibnamefont {Zuo}}, \bibinfo {author} {\bibfnamefont {M.}~\bibnamefont
  {Zhang}}, \bibinfo {author} {\bibfnamefont {Z.}~\bibnamefont {Wei}}, \bibinfo
  {author} {\bibfnamefont {W.}~\bibnamefont {Niu}}, \bibinfo {author}
  {\bibfnamefont {Z.}~\bibnamefont {Xia}}, \bibinfo {author} {\bibfnamefont
  {X.}~\bibnamefont {Wan}}, \bibinfo {author} {\bibfnamefont {Y.}~\bibnamefont
  {Chen}}, \bibinfo {author} {\bibfnamefont {F.}~\bibnamefont {Song}}, \bibinfo
  {author} {\bibfnamefont {Y.}~\bibnamefont {Xu}}, \bibinfo {author}
  {\bibfnamefont {B.}~\bibnamefont {Wang}}, \bibinfo {author} {\bibfnamefont
  {G.}~\bibnamefont {Wang}},\ and\ \bibinfo {author} {\bibfnamefont
  {R.}~\bibnamefont {Zhang}},\ }\bibfield  {title} {\bibinfo {title}
  {\textit{Evidence of Both Surface and Bulk Dirac Bands and Anisotropic
  Nonsaturating Magnetoresistance in ZrSiS}},\ }\href
  {https://doi.org/10.1002/aelm.201600228} {\bibfield  {journal} {\bibinfo
  {journal} {Adv. Electron. Mater}\ }\textbf {\bibinfo {volume} {2}},\ \bibinfo
  {pages} {1600228} (\bibinfo {year} {2016})}\BibitemShut {NoStop}%
\bibitem [{\citenamefont {Chen}\ \emph {et~al.}(2017)\citenamefont {Chen},
  \citenamefont {Xu}, \citenamefont {Jiang}, \citenamefont {Wu}, \citenamefont
  {Qi}, \citenamefont {Yang}, \citenamefont {Wang}, \citenamefont {Sun},
  \citenamefont {Schr{\"o}ter}, \citenamefont {Yang}, \citenamefont {Schoop},
  \citenamefont {Lv}, \citenamefont {Zhou}, \citenamefont {Chen}, \citenamefont
  {Yao}, \citenamefont {Lu}, \citenamefont {Chen}, \citenamefont {Felser},
  \citenamefont {Yan}, \citenamefont {Liu},\ and\ \citenamefont
  {Chen}}]{Chen.2017}%
  \BibitemOpen
  \bibfield  {author} {\bibinfo {author} {\bibfnamefont {C.}~\bibnamefont
  {Chen}}, \bibinfo {author} {\bibfnamefont {X.}~\bibnamefont {Xu}}, \bibinfo
  {author} {\bibfnamefont {J.}~\bibnamefont {Jiang}}, \bibinfo {author}
  {\bibfnamefont {S.-C.}\ \bibnamefont {Wu}}, \bibinfo {author} {\bibfnamefont
  {Y.~P.}\ \bibnamefont {Qi}}, \bibinfo {author} {\bibfnamefont {L.~X.}\
  \bibnamefont {Yang}}, \bibinfo {author} {\bibfnamefont {M.~X.}\ \bibnamefont
  {Wang}}, \bibinfo {author} {\bibfnamefont {Y.}~\bibnamefont {Sun}}, \bibinfo
  {author} {\bibfnamefont {N.~B.~M.}\ \bibnamefont {Schr{\"o}ter}}, \bibinfo
  {author} {\bibfnamefont {H.~F.}\ \bibnamefont {Yang}}, \bibinfo {author}
  {\bibfnamefont {L.~M.}\ \bibnamefont {Schoop}}, \bibinfo {author}
  {\bibfnamefont {Y.~Y.}\ \bibnamefont {Lv}}, \bibinfo {author} {\bibfnamefont
  {J.}~\bibnamefont {Zhou}}, \bibinfo {author} {\bibfnamefont {Y.~B.}\
  \bibnamefont {Chen}}, \bibinfo {author} {\bibfnamefont {S.~H.}\ \bibnamefont
  {Yao}}, \bibinfo {author} {\bibfnamefont {M.~H.}\ \bibnamefont {Lu}},
  \bibinfo {author} {\bibfnamefont {Y.~F.}\ \bibnamefont {Chen}}, \bibinfo
  {author} {\bibfnamefont {C.}~\bibnamefont {Felser}}, \bibinfo {author}
  {\bibfnamefont {B.~H.}\ \bibnamefont {Yan}}, \bibinfo {author} {\bibfnamefont
  {Z.~K.}\ \bibnamefont {Liu}},\ and\ \bibinfo {author} {\bibfnamefont {Y.~L.}\
  \bibnamefont {Chen}},\ }\bibfield  {title} {\bibinfo {title} {\textit{Dirac
  Line Nodes and Effect of Spin-orbit Coupling in the Nonsymmorphic Critical
  Semimetals MSiS(M=Hf,Zr)}},\ }\href
  {https://doi.org/10.1103/PhysRevB.95.125126} {\bibfield  {journal} {\bibinfo
  {journal} {Phys. Rev. B}\ }\textbf {\bibinfo {volume} {95}},\ \bibinfo
  {pages} {125126} (\bibinfo {year} {2017})}\BibitemShut {NoStop}%
\bibitem [{\citenamefont {Hosen}\ \emph {et~al.}(2017)\citenamefont {Hosen},
  \citenamefont {Dimitri}, \citenamefont {Belopolski}, \citenamefont
  {Maldonado}, \citenamefont {Sankar}, \citenamefont {Dhakal}, \citenamefont
  {Dhakal}, \citenamefont {Cole}, \citenamefont {Oppeneer}, \citenamefont
  {Kaczorowski}, \citenamefont {Chou}, \citenamefont {Hasan}, \citenamefont
  {Durakiewicz},\ and\ \citenamefont {Neupane}}]{Hosen.2017}%
  \BibitemOpen
  \bibfield  {author} {\bibinfo {author} {\bibfnamefont {M.~M.}\ \bibnamefont
  {Hosen}}, \bibinfo {author} {\bibfnamefont {K.}~\bibnamefont {Dimitri}},
  \bibinfo {author} {\bibfnamefont {I.}~\bibnamefont {Belopolski}}, \bibinfo
  {author} {\bibfnamefont {P.}~\bibnamefont {Maldonado}}, \bibinfo {author}
  {\bibfnamefont {R.}~\bibnamefont {Sankar}}, \bibinfo {author} {\bibfnamefont
  {N.}~\bibnamefont {Dhakal}}, \bibinfo {author} {\bibfnamefont
  {G.}~\bibnamefont {Dhakal}}, \bibinfo {author} {\bibfnamefont
  {T.}~\bibnamefont {Cole}}, \bibinfo {author} {\bibfnamefont {P.~M.}\
  \bibnamefont {Oppeneer}}, \bibinfo {author} {\bibfnamefont {D.}~\bibnamefont
  {Kaczorowski}}, \bibinfo {author} {\bibfnamefont {F.}~\bibnamefont {Chou}},
  \bibinfo {author} {\bibfnamefont {M.~Z.}\ \bibnamefont {Hasan}}, \bibinfo
  {author} {\bibfnamefont {T.}~\bibnamefont {Durakiewicz}},\ and\ \bibinfo
  {author} {\bibfnamefont {M.}~\bibnamefont {Neupane}},\ }\bibfield  {title}
  {\bibinfo {title} {\textit{Tunability of the Topological Nodal-line Semimetal
  Phase in ${ ZrSiX}$-type Materials ( X=S, Se, Te )}},\ }\href
  {https://doi.org/10.1103/PhysRevB.95.161101} {\bibfield  {journal} {\bibinfo
  {journal} {Phys. Rev. B}\ }\textbf {\bibinfo {volume} {95}},\ \bibinfo
  {pages} {161101(R)} (\bibinfo {year} {2017})}\BibitemShut {NoStop}%
\bibitem [{\citenamefont {Fu}\ \emph {et~al.}(2019)\citenamefont {Fu},
  \citenamefont {Yi}, \citenamefont {Zhang}, \citenamefont {Caputo},
  \citenamefont {Ma}, \citenamefont {Gao}, \citenamefont {Lv}, \citenamefont
  {Kong}, \citenamefont {Huang}, \citenamefont {Richard}, \citenamefont {Shi},
  \citenamefont {Strocov}, \citenamefont {Fang}, \citenamefont {Weng},
  \citenamefont {Shi}, \citenamefont {Qian},\ and\ \citenamefont
  {Ding}}]{Fu.2019}%
  \BibitemOpen
  \bibfield  {author} {\bibinfo {author} {\bibfnamefont {B.-B.}\ \bibnamefont
  {Fu}}, \bibinfo {author} {\bibfnamefont {C.-J.}\ \bibnamefont {Yi}}, \bibinfo
  {author} {\bibfnamefont {T.-T.}\ \bibnamefont {Zhang}}, \bibinfo {author}
  {\bibfnamefont {M.}~\bibnamefont {Caputo}}, \bibinfo {author} {\bibfnamefont
  {J.-Z.}\ \bibnamefont {Ma}}, \bibinfo {author} {\bibfnamefont
  {X.}~\bibnamefont {Gao}}, \bibinfo {author} {\bibfnamefont {B.~Q.}\
  \bibnamefont {Lv}}, \bibinfo {author} {\bibfnamefont {L.-Y.}\ \bibnamefont
  {Kong}}, \bibinfo {author} {\bibfnamefont {Y.-B.}\ \bibnamefont {Huang}},
  \bibinfo {author} {\bibfnamefont {P.}~\bibnamefont {Richard}}, \bibinfo
  {author} {\bibfnamefont {M.}~\bibnamefont {Shi}}, \bibinfo {author}
  {\bibfnamefont {V.~N.}\ \bibnamefont {Strocov}}, \bibinfo {author}
  {\bibfnamefont {C.}~\bibnamefont {Fang}}, \bibinfo {author} {\bibfnamefont
  {H.-M.}\ \bibnamefont {Weng}}, \bibinfo {author} {\bibfnamefont {Y.-G.}\
  \bibnamefont {Shi}}, \bibinfo {author} {\bibfnamefont {T.}~\bibnamefont
  {Qian}},\ and\ \bibinfo {author} {\bibfnamefont {H.}~\bibnamefont {Ding}},\
  }\bibfield  {title} {\bibinfo {title} {\textit{Dirac Nodal Surfaces and Nodal
  Lines in ZrSiS}},\ }\href {https://doi.org/10.1126/sciadv.aau6459} {\bibfield
   {journal} {\bibinfo  {journal} {Sci. Adv.}\ }\textbf {\bibinfo {volume}
  {5}},\ \bibinfo {pages} {eaau6459} (\bibinfo {year} {2019})}\BibitemShut
  {NoStop}%
\bibitem [{\citenamefont {Ali}\ \emph {et~al.}(2016)\citenamefont {Ali},
  \citenamefont {Schoop}, \citenamefont {Garg}, \citenamefont {Lippmann},
  \citenamefont {Lara}, \citenamefont {Lotsch},\ and\ \citenamefont
  {Parkin}}]{Ali.2016}%
  \BibitemOpen
  \bibfield  {author} {\bibinfo {author} {\bibfnamefont {M.~N.}\ \bibnamefont
  {Ali}}, \bibinfo {author} {\bibfnamefont {L.~M.}\ \bibnamefont {Schoop}},
  \bibinfo {author} {\bibfnamefont {C.}~\bibnamefont {Garg}}, \bibinfo {author}
  {\bibfnamefont {J.~M.}\ \bibnamefont {Lippmann}}, \bibinfo {author}
  {\bibfnamefont {E.}~\bibnamefont {Lara}}, \bibinfo {author} {\bibfnamefont
  {B.}~\bibnamefont {Lotsch}},\ and\ \bibinfo {author} {\bibfnamefont
  {S.~S.~P.}\ \bibnamefont {Parkin}},\ }\bibfield  {title} {\bibinfo {title}
  {\textit{Butterfly Magnetoresistance, Quasi-2D Dirac Fermi Surface and
  Topological Phase Transition in ZrSiS}},\ }\href
  {https://doi.org/10.1126/sciadv.1601742} {\bibfield  {journal} {\bibinfo
  {journal} {Sci. Adv.}\ }\textbf {\bibinfo {volume} {2}},\ \bibinfo {pages}
  {e1601742} (\bibinfo {year} {2016})}\BibitemShut {NoStop}%
\bibitem [{\citenamefont {Lv}\ \emph {et~al.}(2016)\citenamefont {Lv},
  \citenamefont {Zhang}, \citenamefont {Li}, \citenamefont {Yao}, \citenamefont
  {Chen}, \citenamefont {Zhou}, \citenamefont {Zhang}, \citenamefont {Lu},\
  and\ \citenamefont {Chen}}]{Lv.2016}%
  \BibitemOpen
  \bibfield  {author} {\bibinfo {author} {\bibfnamefont {Y.-Y.}\ \bibnamefont
  {Lv}}, \bibinfo {author} {\bibfnamefont {B.-B.}\ \bibnamefont {Zhang}},
  \bibinfo {author} {\bibfnamefont {X.}~\bibnamefont {Li}}, \bibinfo {author}
  {\bibfnamefont {S.-H.}\ \bibnamefont {Yao}}, \bibinfo {author} {\bibfnamefont
  {Y.~B.}\ \bibnamefont {Chen}}, \bibinfo {author} {\bibfnamefont
  {J.}~\bibnamefont {Zhou}}, \bibinfo {author} {\bibfnamefont {S.-T.}\
  \bibnamefont {Zhang}}, \bibinfo {author} {\bibfnamefont {M.-H.}\ \bibnamefont
  {Lu}},\ and\ \bibinfo {author} {\bibfnamefont {Y.-F.}\ \bibnamefont {Chen}},\
  }\bibfield  {title} {\bibinfo {title} {\textit{Extremely Large and
  Significantly Anisotropic Magnetoresistance in ZrSiS Single Crystals}},\
  }\href {https://doi.org/10.1063/1.4953772} {\bibfield  {journal} {\bibinfo
  {journal} {Appl. Phys. Lett.}\ }\textbf {\bibinfo {volume} {108}},\ \bibinfo
  {pages} {244101} (\bibinfo {year} {2016})}\BibitemShut {NoStop}%
\bibitem [{\citenamefont {Singha}\ \emph {et~al.}(2017)\citenamefont {Singha},
  \citenamefont {Pariari}, \citenamefont {Satpati},\ and\ \citenamefont
  {Mandal}}]{Singha.2017}%
  \BibitemOpen
  \bibfield  {author} {\bibinfo {author} {\bibfnamefont {R.}~\bibnamefont
  {Singha}}, \bibinfo {author} {\bibfnamefont {A.~K.}\ \bibnamefont {Pariari}},
  \bibinfo {author} {\bibfnamefont {B.}~\bibnamefont {Satpati}},\ and\ \bibinfo
  {author} {\bibfnamefont {P.}~\bibnamefont {Mandal}},\ }\bibfield  {title}
  {\bibinfo {title} {\textit{Large Nonsaturating Magnetoresistance and
  Signature of Nondegenerate Dirac Nodes in ZrSiS}},\ }\href
  {https://doi.org/10.1073/pnas.1618004114} {\bibfield  {journal} {\bibinfo
  {journal} {Proc. Natl. Acad. Sci. U.S.A.}\ }\textbf {\bibinfo {volume}
  {114}},\ \bibinfo {pages} {2468} (\bibinfo {year} {2017})}\BibitemShut
  {NoStop}%
\bibitem [{\citenamefont {Hu}\ \emph {et~al.}(2017)\citenamefont {Hu},
  \citenamefont {Tang}, \citenamefont {Liu}, \citenamefont {Zhu}, \citenamefont
  {Wei},\ and\ \citenamefont {Mao}}]{Hu.2017}%
  \BibitemOpen
  \bibfield  {author} {\bibinfo {author} {\bibfnamefont {J.}~\bibnamefont
  {Hu}}, \bibinfo {author} {\bibfnamefont {Z.}~\bibnamefont {Tang}}, \bibinfo
  {author} {\bibfnamefont {J.}~\bibnamefont {Liu}}, \bibinfo {author}
  {\bibfnamefont {Y.}~\bibnamefont {Zhu}}, \bibinfo {author} {\bibfnamefont
  {J.}~\bibnamefont {Wei}},\ and\ \bibinfo {author} {\bibfnamefont
  {Z.}~\bibnamefont {Mao}},\ }\bibfield  {title} {\bibinfo {title}
  {\textit{Nearly Massless Dirac Fermions and Strong Zeeman Splitting in the
  Nodal-line Semimetal ZrSiS Probed by de Haas-van Alphen Quantum
  Oscillations}},\ }\href {https://doi.org/10.1103/PhysRevB.96.045127}
  {\bibfield  {journal} {\bibinfo  {journal} {Phys. Rev. B}\ }\textbf {\bibinfo
  {volume} {96}},\ \bibinfo {pages} {045127} (\bibinfo {year}
  {2017})}\BibitemShut {NoStop}%
\bibitem [{\citenamefont {Pezzini}\ \emph {et~al.}(2018)\citenamefont
  {Pezzini}, \citenamefont {{van Delft}}, \citenamefont {Schoop}, \citenamefont
  {Lotsch}, \citenamefont {Carrington}, \citenamefont {Katsnelson},
  \citenamefont {Hussey},\ and\ \citenamefont {Wiedmann}}]{Pezzini.2018}%
  \BibitemOpen
  \bibfield  {author} {\bibinfo {author} {\bibfnamefont {S.}~\bibnamefont
  {Pezzini}}, \bibinfo {author} {\bibfnamefont {M.~R.}\ \bibnamefont {{van
  Delft}}}, \bibinfo {author} {\bibfnamefont {L.~M.}\ \bibnamefont {Schoop}},
  \bibinfo {author} {\bibfnamefont {B.~V.}\ \bibnamefont {Lotsch}}, \bibinfo
  {author} {\bibfnamefont {A.}~\bibnamefont {Carrington}}, \bibinfo {author}
  {\bibfnamefont {M.~I.}\ \bibnamefont {Katsnelson}}, \bibinfo {author}
  {\bibfnamefont {N.~E.}\ \bibnamefont {Hussey}},\ and\ \bibinfo {author}
  {\bibfnamefont {S.}~\bibnamefont {Wiedmann}},\ }\bibfield  {title} {\bibinfo
  {title} {\textit{Unconventional Mass Enhancement Around the Dirac Nodal Loop
  in Zrsis}},\ }\href {https://doi.org/10.1038/NPHYS4306} {\bibfield  {journal}
  {\bibinfo  {journal} {Nat. Phys.}\ }\textbf {\bibinfo {volume} {14}},\
  \bibinfo {pages} {178} (\bibinfo {year} {2018})}\BibitemShut {NoStop}%
\bibitem [{\citenamefont {Zhu}\ \emph {et~al.}(2015)\citenamefont {Zhu},
  \citenamefont {Cao}, \citenamefont {Zhang}, \citenamefont {Jia},
  \citenamefont {Guo}, \citenamefont {Yang}, \citenamefont {Zhu}, \citenamefont
  {Zhang}, \citenamefont {Plummer},\ and\ \citenamefont {Guo}}]{Zhu.2015}%
  \BibitemOpen
  \bibfield  {author} {\bibinfo {author} {\bibfnamefont {X.}~\bibnamefont
  {Zhu}}, \bibinfo {author} {\bibfnamefont {Y.}~\bibnamefont {Cao}}, \bibinfo
  {author} {\bibfnamefont {S.}~\bibnamefont {Zhang}}, \bibinfo {author}
  {\bibfnamefont {X.}~\bibnamefont {Jia}}, \bibinfo {author} {\bibfnamefont
  {Q.}~\bibnamefont {Guo}}, \bibinfo {author} {\bibfnamefont {F.}~\bibnamefont
  {Yang}}, \bibinfo {author} {\bibfnamefont {L.}~\bibnamefont {Zhu}}, \bibinfo
  {author} {\bibfnamefont {J.}~\bibnamefont {Zhang}}, \bibinfo {author}
  {\bibfnamefont {E.~W.}\ \bibnamefont {Plummer}},\ and\ \bibinfo {author}
  {\bibfnamefont {J.}~\bibnamefont {Guo}},\ }\bibfield  {title} {\bibinfo
  {title} {\textit{High Resolution Electron Energy Loss Spectroscopy with
  Two-dimensional Energy and Momentum Mapping}},\ }\href
  {https://doi.org/10.1063/1.4928215} {\bibfield  {journal} {\bibinfo
  {journal} {Rev. Sci. Instrum.}\ }\textbf {\bibinfo {volume} {86}},\ \bibinfo
  {pages} {083902} (\bibinfo {year} {2015})}\BibitemShut {NoStop}%
\bibitem [{\citenamefont {Kresse}\ and\ \citenamefont
  {Furthm{\"u}ller}(1996)}]{Kresse.1996}%
  \BibitemOpen
  \bibfield  {author} {\bibinfo {author} {\bibnamefont {Kresse}}\ and\ \bibinfo
  {author} {\bibnamefont {Furthm{\"u}ller}},\ }\bibfield  {title} {\bibinfo
  {title} {\textit{Efficient Iterative Schemes for ab Initio Total-energy
  Calculations Using a Plane-wave Basis Set}},\ }\href
  {https://doi.org/10.1103/PhysRevB.54.11169} {\bibfield  {journal} {\bibinfo
  {journal} {Phys. Rev. B}\ }\textbf {\bibinfo {volume} {54}},\ \bibinfo
  {pages} {11169} (\bibinfo {year} {1996})}\BibitemShut {NoStop}%
\bibitem [{\citenamefont {Baroni}\ \emph {et~al.}(2001)\citenamefont {Baroni},
  \citenamefont {de~Gironcoli}, \citenamefont {{Dal Corso}},\ and\
  \citenamefont {Giannozzi}}]{Baroni.2001}%
  \BibitemOpen
  \bibfield  {author} {\bibinfo {author} {\bibfnamefont {S.}~\bibnamefont
  {Baroni}}, \bibinfo {author} {\bibfnamefont {S.}~\bibnamefont
  {de~Gironcoli}}, \bibinfo {author} {\bibfnamefont {A.}~\bibnamefont {{Dal
  Corso}}},\ and\ \bibinfo {author} {\bibfnamefont {P.}~\bibnamefont
  {Giannozzi}},\ }\bibfield  {title} {\bibinfo {title} {\textit{Phonons And
  Related Crystal Properties From Density-functional Perturbation Theory}},\
  }\href {https://doi.org/10.1103/RevModPhys.73.515} {\bibfield  {journal}
  {\bibinfo  {journal} {Rev. Mod. Phys.}\ }\textbf {\bibinfo {volume} {73}},\
  \bibinfo {pages} {515} (\bibinfo {year} {2001})}\BibitemShut {NoStop}%
\bibitem [{\citenamefont {Perdew}\ \emph {et~al.}(1992)\citenamefont {Perdew},
  \citenamefont {Chevary}, \citenamefont {Vosko}, \citenamefont {Jackson},
  \citenamefont {Pederson}, \citenamefont {Singh},\ and\ \citenamefont
  {Fiolhais}}]{Perdew.1992}%
  \BibitemOpen
  \bibfield  {author} {\bibinfo {author} {\bibnamefont {Perdew}}, \bibinfo
  {author} {\bibnamefont {Chevary}}, \bibinfo {author} {\bibnamefont {Vosko}},
  \bibinfo {author} {\bibnamefont {Jackson}}, \bibinfo {author} {\bibnamefont
  {Pederson}}, \bibinfo {author} {\bibnamefont {Singh}},\ and\ \bibinfo
  {author} {\bibnamefont {Fiolhais}},\ }\bibfield  {title} {\bibinfo {title}
  {\textit{Atoms, Molecules, Solids, and Surfaces: Applications of the
  Generalized Gradient Approximation for Exchange and Correlation}},\ }\href
  {https://doi.org/10.1103/PhysRevB.46.6671} {\bibfield  {journal} {\bibinfo
  {journal} {Phys. Rev. B}\ }\textbf {\bibinfo {volume} {46}},\ \bibinfo
  {pages} {6671} (\bibinfo {year} {1992})}\BibitemShut {NoStop}%
\bibitem [{\citenamefont {Bl{\"o}chl}(1994)}]{Blochl.1994}%
  \BibitemOpen
  \bibfield  {author} {\bibinfo {author} {\bibnamefont {Bl{\"o}chl}},\
  }\bibfield  {title} {\bibinfo {title} {\textit{Projector Augmented-Wave
  Method}},\ }\href {https://doi.org/10.1103/PhysRevB.50.17953} {\bibfield
  {journal} {\bibinfo  {journal} {Phys. Rev. B}\ }\textbf {\bibinfo {volume}
  {50}},\ \bibinfo {pages} {17953} (\bibinfo {year} {1994})}\BibitemShut
  {NoStop}%
\bibitem [{\citenamefont {Togo}\ and\ \citenamefont
  {Tanaka}(2015)}]{Togo.2015}%
  \BibitemOpen
  \bibfield  {author} {\bibinfo {author} {\bibfnamefont {A.}~\bibnamefont
  {Togo}}\ and\ \bibinfo {author} {\bibfnamefont {I.}~\bibnamefont {Tanaka}},\
  }\bibfield  {title} {\bibinfo {title} {\textit{First Principles Phonon
  Calculations in Materials Science}},\ }\href
  {https://doi.org/10.1016/j.scriptamat.2015.07.021} {\bibfield  {journal}
  {\bibinfo  {journal} {Scr. Mater.}\ }\textbf {\bibinfo {volume} {108}},\
  \bibinfo {pages} {1} (\bibinfo {year} {2015})}\BibitemShut {NoStop}%
\bibitem [{\citenamefont {Wu}\ \emph {et~al.}(2018)\citenamefont {Wu},
  \citenamefont {Zhang}, \citenamefont {Song}, \citenamefont {Troyer},\ and\
  \citenamefont {Soluyanov}}]{Wu.2018}%
  \BibitemOpen
  \bibfield  {author} {\bibinfo {author} {\bibfnamefont {Q.}~\bibnamefont
  {Wu}}, \bibinfo {author} {\bibfnamefont {S.}~\bibnamefont {Zhang}}, \bibinfo
  {author} {\bibfnamefont {H.-F.}\ \bibnamefont {Song}}, \bibinfo {author}
  {\bibfnamefont {M.}~\bibnamefont {Troyer}},\ and\ \bibinfo {author}
  {\bibfnamefont {A.~A.}\ \bibnamefont {Soluyanov}},\ }\bibfield  {title}
  {\bibinfo {title} {\textit{WannierTools: An open-source Software Package for
  Novel Topological Materials}},\ }\href
  {https://doi.org/10.1016/j.cpc.2017.09.033} {\bibfield  {journal} {\bibinfo
  {journal} {Comput. Phys. Commun.}\ }\textbf {\bibinfo {volume} {224}},\
  \bibinfo {pages} {405} (\bibinfo {year} {2018})}\BibitemShut {NoStop}%
\bibitem [{\citenamefont {{Klein Haneveld}}\ and\ \citenamefont
  {Jellinek}(1964)}]{KleinHaneveld.1964}%
  \BibitemOpen
  \bibfield  {author} {\bibinfo {author} {\bibfnamefont {A.~J.}\ \bibnamefont
  {{Klein Haneveld}}}\ and\ \bibinfo {author} {\bibfnamefont {F.}~\bibnamefont
  {Jellinek}},\ }\bibfield  {title} {\bibinfo {title} {\textit{Zirconium
  Silicide and Germanide Chalcogenides Preparation and Crystal Structures}},\
  }\href {https://doi.org/10.1002/recl.19640830802} {\bibfield  {journal}
  {\bibinfo  {journal} {Recl. Trav. Chim. Pays-Bas}\ }\textbf {\bibinfo
  {volume} {83}},\ \bibinfo {pages} {776} (\bibinfo {year} {1964})}\BibitemShut
  {NoStop}%
\bibitem [{\citenamefont {Tremel}\ and\ \citenamefont
  {Hoffmann}(1987)}]{Tremel.1987}%
  \BibitemOpen
  \bibfield  {author} {\bibinfo {author} {\bibfnamefont {W.}~\bibnamefont
  {Tremel}}\ and\ \bibinfo {author} {\bibfnamefont {R.}~\bibnamefont
  {Hoffmann}},\ }\bibfield  {title} {\bibinfo {title} {\textit{Square Nets of
  Main-group Elements in Solid-state Materials}},\ }\href
  {https://doi.org/10.1021/ja00235a021} {\bibfield  {journal} {\bibinfo
  {journal} {J. Am. Chem. Soc.}\ }\textbf {\bibinfo {volume} {109}},\ \bibinfo
  {pages} {124} (\bibinfo {year} {1987})}\BibitemShut {NoStop}%
\bibitem [{\citenamefont {Singha}\ \emph {et~al.}(2018)\citenamefont {Singha},
  \citenamefont {Samanta}, \citenamefont {Chatterjee}, \citenamefont {Pariari},
  \citenamefont {Majumdar}, \citenamefont {Satpati}, \citenamefont {Wang},
  \citenamefont {Singha},\ and\ \citenamefont {Mandal}}]{Singha.2018}%
  \BibitemOpen
  \bibfield  {author} {\bibinfo {author} {\bibfnamefont {R.}~\bibnamefont
  {Singha}}, \bibinfo {author} {\bibfnamefont {S.}~\bibnamefont {Samanta}},
  \bibinfo {author} {\bibfnamefont {S.}~\bibnamefont {Chatterjee}}, \bibinfo
  {author} {\bibfnamefont {A.}~\bibnamefont {Pariari}}, \bibinfo {author}
  {\bibfnamefont {D.}~\bibnamefont {Majumdar}}, \bibinfo {author}
  {\bibfnamefont {B.}~\bibnamefont {Satpati}}, \bibinfo {author} {\bibfnamefont
  {L.}~\bibnamefont {Wang}}, \bibinfo {author} {\bibfnamefont {A.}~\bibnamefont
  {Singha}},\ and\ \bibinfo {author} {\bibfnamefont {P.}~\bibnamefont
  {Mandal}},\ }\bibfield  {title} {\bibinfo {title} {\textit{Probing Lattice
  Dynamics and Electron-phonon Coupling in the Topological Nodal-line Semimetal
  ZrSiS}},\ }\href {https://doi.org/10.1103/PhysRevB.97.094112} {\bibfield
  {journal} {\bibinfo  {journal} {Phys. Rev. B}\ }\textbf {\bibinfo {volume}
  {97}},\ \bibinfo {pages} {094112} (\bibinfo {year} {2018})}\BibitemShut
  {NoStop}%
\bibitem [{\citenamefont {Salmankurt}\ and\ \citenamefont
  {Duman}(2017)}]{Salmankurt.2017}%
  \BibitemOpen
  \bibfield  {author} {\bibinfo {author} {\bibfnamefont {B.}~\bibnamefont
  {Salmankurt}}\ and\ \bibinfo {author} {\bibfnamefont {S.}~\bibnamefont
  {Duman}},\ }\bibfield  {title} {\bibinfo {title} {\textit{First-principles
  Study of Structural, Mechanical, Lattice Dynamical and Thermal Properties of
  Nodal-line Semimetals ZrXY (X=Si,Ge; Y=S,Se)}},\ }\href
  {https://doi.org/10.1080/14786435.2016.1250967} {\bibfield  {journal}
  {\bibinfo  {journal} {Philos. Mag.}\ }\textbf {\bibinfo {volume} {97}},\
  \bibinfo {pages} {175} (\bibinfo {year} {2017})}\BibitemShut {NoStop}%
\bibitem [{\citenamefont {Zhou}\ \emph {et~al.}(2017)\citenamefont {Zhou},
  \citenamefont {Gao}, \citenamefont {Zhang}, \citenamefont {Fang},
  \citenamefont {Song}, \citenamefont {Hu}, \citenamefont {Stroppa},
  \citenamefont {Li}, \citenamefont {Wang}, \citenamefont {Ruan},\ and\
  \citenamefont {Ren}}]{Zhou.2017}%
  \BibitemOpen
  \bibfield  {author} {\bibinfo {author} {\bibfnamefont {W.}~\bibnamefont
  {Zhou}}, \bibinfo {author} {\bibfnamefont {H.}~\bibnamefont {Gao}}, \bibinfo
  {author} {\bibfnamefont {J.}~\bibnamefont {Zhang}}, \bibinfo {author}
  {\bibfnamefont {R.}~\bibnamefont {Fang}}, \bibinfo {author} {\bibfnamefont
  {H.}~\bibnamefont {Song}}, \bibinfo {author} {\bibfnamefont {T.}~\bibnamefont
  {Hu}}, \bibinfo {author} {\bibfnamefont {A.}~\bibnamefont {Stroppa}},
  \bibinfo {author} {\bibfnamefont {L.}~\bibnamefont {Li}}, \bibinfo {author}
  {\bibfnamefont {X.}~\bibnamefont {Wang}}, \bibinfo {author} {\bibfnamefont
  {S.}~\bibnamefont {Ruan}},\ and\ \bibinfo {author} {\bibfnamefont
  {W.}~\bibnamefont {Ren}},\ }\bibfield  {title} {\bibinfo {title}
  {\textit{Lattice Dynamics of Dirac Node-line Semimetal ZrSiS}},\ }\href
  {https://doi.org/10.1103/PhysRevB.96.064103} {\bibfield  {journal} {\bibinfo
  {journal} {Phys. Rev. B}\ }\textbf {\bibinfo {volume} {96}},\ \bibinfo
  {pages} {064103} (\bibinfo {year} {2017})}\BibitemShut {NoStop}%
\bibitem [{\citenamefont {Ibach}\ and\ \citenamefont
  {Mills}(1982)}]{ibach.1982}%
  \BibitemOpen
  \bibfield  {author} {\bibinfo {author} {\bibfnamefont {H.}~\bibnamefont
  {Ibach}}\ and\ \bibinfo {author} {\bibfnamefont {D.~L.}\ \bibnamefont
  {Mills}},\ }\href@noop {} {\emph {\bibinfo {title} {Electron energy loss
  spectroscopy and surface vibrations}}}\ (\bibinfo  {publisher} {Academic
  Press New York},\ \bibinfo {year} {1982})\BibitemShut {NoStop}%
\bibitem [{\citenamefont {de~Juan}\ \emph {et~al.}(2015)\citenamefont
  {de~Juan}, \citenamefont {Politano}, \citenamefont {Chiarello},\ and\
  \citenamefont {Fertig}}]{Juan.2015}%
  \BibitemOpen
  \bibfield  {author} {\bibinfo {author} {\bibfnamefont {F.}~\bibnamefont
  {de~Juan}}, \bibinfo {author} {\bibfnamefont {A.}~\bibnamefont {Politano}},
  \bibinfo {author} {\bibfnamefont {G.}~\bibnamefont {Chiarello}},\ and\
  \bibinfo {author} {\bibfnamefont {H.~A.}\ \bibnamefont {Fertig}},\ }\bibfield
   {title} {\bibinfo {title} {\textit{Symmetries and Selection Rules in the
  Measurement of the Phonon Spectrum of Graphene and Related Materials}},\
  }\href {https://doi.org/10.1016/j.carbon.2014.12.105} {\bibfield  {journal}
  {\bibinfo  {journal} {Carbon}\ }\textbf {\bibinfo {volume} {85}},\ \bibinfo
  {pages} {225} (\bibinfo {year} {2015})}\BibitemShut {NoStop}%
\bibitem [{\citenamefont {Huh}\ \emph {et~al.}(2016)\citenamefont {Huh},
  \citenamefont {Moon},\ and\ \citenamefont {Kim}}]{Huh.2016}%
  \BibitemOpen
  \bibfield  {author} {\bibinfo {author} {\bibfnamefont {Y.}~\bibnamefont
  {Huh}}, \bibinfo {author} {\bibfnamefont {E.-G.}\ \bibnamefont {Moon}},\ and\
  \bibinfo {author} {\bibfnamefont {Y.~B.}\ \bibnamefont {Kim}},\ }\bibfield
  {title} {\bibinfo {title} {\textit{Long-range Coulomb interaction in
  Nodal-ring Semimetals}},\ }\href {https://doi.org/10.1103/PhysRevB.93.035138}
  {\bibfield  {journal} {\bibinfo  {journal} {Phys. Rev. B}\ }\textbf {\bibinfo
  {volume} {93}},\ \bibinfo {pages} {035138} (\bibinfo {year}
  {2016})}\BibitemShut {NoStop}%
\bibitem [{\citenamefont {Roy}(2017)}]{Roy.2017}%
  \BibitemOpen
  \bibfield  {author} {\bibinfo {author} {\bibfnamefont {B.}~\bibnamefont
  {Roy}},\ }\bibfield  {title} {\bibinfo {title} {\textit{Interacting
  Nodal-line Semimetal: Proximity Effect and Spontaneous Symmetry Breaking}},\
  }\href {https://doi.org/10.1103/PhysRevB.96.041113} {\bibfield  {journal}
  {\bibinfo  {journal} {Phys. Rev. B}\ }\textbf {\bibinfo {volume} {96}},\
  \bibinfo {pages} {041113(R)} (\bibinfo {year} {2017})}\BibitemShut {NoStop}%
\bibitem [{\citenamefont {Butler}\ \emph {et~al.}(2017)\citenamefont {Butler},
  \citenamefont {Wu}, \citenamefont {Hsing}, \citenamefont {Tseng},
  \citenamefont {Sankar}, \citenamefont {Wei}, \citenamefont {Chou},\ and\
  \citenamefont {Lin}}]{Butler.2017}%
  \BibitemOpen
  \bibfield  {author} {\bibinfo {author} {\bibfnamefont {C.~J.}\ \bibnamefont
  {Butler}}, \bibinfo {author} {\bibfnamefont {Y.-M.}\ \bibnamefont {Wu}},
  \bibinfo {author} {\bibfnamefont {C.-R.}\ \bibnamefont {Hsing}}, \bibinfo
  {author} {\bibfnamefont {Y.}~\bibnamefont {Tseng}}, \bibinfo {author}
  {\bibfnamefont {R.}~\bibnamefont {Sankar}}, \bibinfo {author} {\bibfnamefont
  {C.-M.}\ \bibnamefont {Wei}}, \bibinfo {author} {\bibfnamefont {F.-C.}\
  \bibnamefont {Chou}},\ and\ \bibinfo {author} {\bibfnamefont {M.-T.}\
  \bibnamefont {Lin}},\ }\bibfield  {title} {\bibinfo {title}
  {\textit{Quasiparticle Interference in ZrSiS: Strongly Band-selective
  Scattering Depending on Impurity Lattice Site}},\ }\href
  {https://doi.org/10.1103/PhysRevB.96.195125} {\bibfield  {journal} {\bibinfo
  {journal} {Phys. Rev. B}\ }\textbf {\bibinfo {volume} {96}},\ \bibinfo
  {pages} {195125} (\bibinfo {year} {2017})}\BibitemShut {NoStop}%
\bibitem [{\citenamefont {Su}\ \emph {et~al.}(2018)\citenamefont {Su},
  \citenamefont {Li}, \citenamefont {Wang}, \citenamefont {Guan}, \citenamefont
  {Sankar}, \citenamefont {Chou}, \citenamefont {Chang}, \citenamefont {Lee},
  \citenamefont {Guo},\ and\ \citenamefont {Chuang}}]{Su.2018}%
  \BibitemOpen
  \bibfield  {author} {\bibinfo {author} {\bibfnamefont {C.-C.}\ \bibnamefont
  {Su}}, \bibinfo {author} {\bibfnamefont {C.-S.}\ \bibnamefont {Li}}, \bibinfo
  {author} {\bibfnamefont {T.-C.}\ \bibnamefont {Wang}}, \bibinfo {author}
  {\bibfnamefont {S.-Y.}\ \bibnamefont {Guan}}, \bibinfo {author}
  {\bibfnamefont {R.}~\bibnamefont {Sankar}}, \bibinfo {author} {\bibfnamefont
  {F.}~\bibnamefont {Chou}}, \bibinfo {author} {\bibfnamefont {C.-S.}\
  \bibnamefont {Chang}}, \bibinfo {author} {\bibfnamefont {W.-L.}\ \bibnamefont
  {Lee}}, \bibinfo {author} {\bibfnamefont {G.-Y.}\ \bibnamefont {Guo}},\ and\
  \bibinfo {author} {\bibfnamefont {T.-M.}\ \bibnamefont {Chuang}},\ }\bibfield
   {title} {\bibinfo {title} {\textit{Surface Termination Dependent
  Quasiparticle Scattering Interference and Magneto-transport Study on
  ZrSiS}},\ }\href {https://doi.org/10.1088/1367-2630/aae5c8} {\bibfield
  {journal} {\bibinfo  {journal} {New J. Phys.}\ }\textbf {\bibinfo {volume}
  {20}},\ \bibinfo {pages} {103025} (\bibinfo {year} {2018})}\BibitemShut
  {NoStop}%
\bibitem [{\citenamefont {Zhu}\ \emph {et~al.}(2018)\citenamefont {Zhu},
  \citenamefont {Chang}, \citenamefont {Huang}, \citenamefont {Pan},
  \citenamefont {Nie}, \citenamefont {Wang}, \citenamefont {Jin}, \citenamefont
  {Xu}, \citenamefont {Huang}, \citenamefont {Guan}, \citenamefont {Wang},
  \citenamefont {Li}, \citenamefont {Liu}, \citenamefont {Qian}, \citenamefont
  {Ku}, \citenamefont {Song}, \citenamefont {Lin}, \citenamefont {Zheng},\ and\
  \citenamefont {Jia}}]{Zhu.2018}%
  \BibitemOpen
  \bibfield  {author} {\bibinfo {author} {\bibfnamefont {Z.}~\bibnamefont
  {Zhu}}, \bibinfo {author} {\bibfnamefont {T.-R.}\ \bibnamefont {Chang}},
  \bibinfo {author} {\bibfnamefont {C.-Y.}\ \bibnamefont {Huang}}, \bibinfo
  {author} {\bibfnamefont {H.}~\bibnamefont {Pan}}, \bibinfo {author}
  {\bibfnamefont {X.-A.}\ \bibnamefont {Nie}}, \bibinfo {author} {\bibfnamefont
  {X.-Z.}\ \bibnamefont {Wang}}, \bibinfo {author} {\bibfnamefont {Z.-T.}\
  \bibnamefont {Jin}}, \bibinfo {author} {\bibfnamefont {S.-Y.}\ \bibnamefont
  {Xu}}, \bibinfo {author} {\bibfnamefont {S.-M.}\ \bibnamefont {Huang}},
  \bibinfo {author} {\bibfnamefont {D.-D.}\ \bibnamefont {Guan}}, \bibinfo
  {author} {\bibfnamefont {S.}~\bibnamefont {Wang}}, \bibinfo {author}
  {\bibfnamefont {Y.-Y.}\ \bibnamefont {Li}}, \bibinfo {author} {\bibfnamefont
  {C.}~\bibnamefont {Liu}}, \bibinfo {author} {\bibfnamefont {D.}~\bibnamefont
  {Qian}}, \bibinfo {author} {\bibfnamefont {W.}~\bibnamefont {Ku}}, \bibinfo
  {author} {\bibfnamefont {F.}~\bibnamefont {Song}}, \bibinfo {author}
  {\bibfnamefont {H.}~\bibnamefont {Lin}}, \bibinfo {author} {\bibfnamefont
  {H.}~\bibnamefont {Zheng}},\ and\ \bibinfo {author} {\bibfnamefont {J.-F.}\
  \bibnamefont {Jia}},\ }\bibfield  {title} {\bibinfo {title}
  {\textit{Quasiparticle Interference and Nonsymmorphic Effect on a Floating
  Band Surface State of ZrSiSe}},\ }\href
  {https://doi.org/10.1038/s41467-018-06661-9} {\bibfield  {journal} {\bibinfo
  {journal} {Nat. Commun.}\ }\textbf {\bibinfo {volume} {9}},\ \bibinfo {pages}
  {4153} (\bibinfo {year} {2018})}\BibitemShut {NoStop}%
\end{thebibliography}%

\end{document}